\documentclass[%
 aip,
 amsmath,amssymb,
 reprint,%
]{revtex4-1}

\usepackage{graphicx}
\usepackage{dcolumn}
\usepackage{bm}

\usepackage[utf8]{inputenc}
\usepackage[T1]{fontenc}
\usepackage{mathptmx}
\usepackage{etoolbox}
\usepackage{bigints}
\usepackage[dvipsnames]{xcolor}
\usepackage{comment} 

\makeatletter
\def\@email#1#2{%
 \endgroup
 \patchcmd{\titleblock@produce}
  {\frontmatter@RRAPformat}
  {\frontmatter@RRAPformat{\produce@RRAP{*#1\href{mailto:#2}{#2}}}\frontmatter@RRAPformat}
  {}{}
}%
\makeatother
\begin{document}

\preprint{AIP/123-QED}

\title[Effect of dispersive interactions in methane and carbon dioxide hydrates]{Three-phase equilibria of hydrates from computer simulation. III. Effect of dispersive interactions in the methane and carbon dioxide hydrates}

\author{J. Algaba$^{\dag}$}
\affiliation{Laboratorio de Simulaci\'on Molecular y Qu\'imica Computacional, CIQSO-Centro de Investigaci\'on en Qu\'imica Sostenible and Departamento de Ciencias Integradas, Universidad de Huelva, 21006 Huelva Spain}

\author{S. Blazquez$^{\dag}$}
\affiliation{Dpto. Qu\'{\i}mica F\'{\i}sica I, Fac. Ciencias Qu\'{\i}micas, Universidad Complutense de Madrid, 28040 Madrid, Spain}

\author{J. M. M\'{\i}guez}
\affiliation{Laboratorio de Simulaci\'on Molecular y Qu\'imica Computacional, CIQSO-Centro de Investigaci\'on en Qu\'imica Sostenible and Departamento de Ciencias Integradas, Universidad de Huelva, 21006 Huelva Spain}

\author{M. M. Conde$^{*}$}
\affiliation{Departamento de Ingeniería Química Industrial y del Medio Ambiente, Escuela Técnica Superior de Ingenieros Industriales, Universidad Politécnica de Madrid,
28006, Madrid, Spain.
}

\author{F. J. Blas$^{*}$}
\affiliation{Laboratorio de Simulaci\'on Molecular y Qu\'imica Computacional, CIQSO-Centro de Investigaci\'on en Qu\'imica Sostenible and Departamento de Ciencias Integradas, Universidad de Huelva, 21006 Huelva Spain}

\begin{abstract}

In this work, the effect of the range of the dispersive interactions in the determination of the three-phase coexistence line of the CO$_2$ and CH$_4$ hydrates has been studied. In particular, the temperature ($T_3$) at which solid hydrate, water, and liquid CO$_2$/gas CH$_4$ coexist has been determined through molecular dynamics simulations using different cut-off values (from 0.9 to 1.6 nm) for the dispersive interactions. The $T_3$ of both hydrates has been determined using the direct coexistence simulation technique. Following this method, the three phases in equilibrium are put together in the same simulation box, the pressure is fixed, and simulations are performed at different temperatures $T$. If the hydrate melts, then $T>T_3$. Contrary, if the hydrate grows, then $T<T_3$. The effect of the cut-off distance on the dissociation temperature has been analyzed at three different pressures for CO$_{2}$ hydrate, namely, $100$, $400$, and $1000\,\text{bar}$. Then, we have changed the guest and studied the effect of the cut-off distance on the dissociation temperature of the CH$_{4}$ hydrate at $400\,\text{bar}$. Also, the effect of long-range corrections for dispersive interactions has been analyzed by running simulations with homo- and inhomogeneous corrections and a cut-off value of 0.9 nm. The results obtained in this work highlight that the cut-off distance for the dispersive interactions affects the stability conditions of these hydrates. This effect is enhanced when the pressure is decreased, displacing the $T_{3}$ about $2-4\,\text{K}$ depending on the system and the pressure.

\end{abstract}

\maketitle
$^*$Corresponding authors: felipe@uhu.es and maria.mconde@upm.es\\
$^{\dag}$These authors contributed equally to this work

%

\section{Introduction}

Clathrate hydrates are non-stochiometric crystalline inclusion compounds consisting of a network of hydrogen-bonded molecules (host) conforming cages in which small molecules (guest) can be encapsulated under the appropriate thermodynamic conditions. When the host molecule is water, these compounds are simply called hydrates.~\cite{Sloan2008a} The structure of the hydrates depends on the thermodynamic conditions, but moreover, depends on the guest molecule encapsulated inside the hydrate. The guest has a high impact on the stability of the hydrate. Hydrates of small molecules, such as methane (CH$_4$) or carbon dioxide (CO$_2$), often crystallize forming the so-called structure sI. Hydrates of medium molecules (i.e., butane, cyclopentane, or tetrahydrofuran) crystallize in structure sII,~\cite{Sloan2008a} although this structure is also stabilized by very small guests such as N$_2$ and H$_2$.~\cite{Barnes2013a} Finally, when there is a mixture of large and small molecules the sH structure is the most common.~\cite{Sloan2008a} 

Due to the capability of hydrates to store CH$_4$\cite{Kvenvolden1988a,Koh2012a} and other low molecular weight hydrocarbons, hydrates are potential sources of natural gas and, therefore, energy. One of the greatest challenges of society is the high energy demand. The fact that the world is constantly evolving towards a more technological society means that energy demand is increasing. Although it is true that more renewable and/or alternative energy sources are being used, in 2022 the main source of energy was fossil fuels (more than 62\% of the electricity generated in the world).~\cite{WEO2022} It is estimated that the amount of natural gas stored in hydrates far exceeds the amount of natural gas available from conventional sources. Other interesting applications of hydrates from a social, environmental, and industrial point of view is their capability to sequester CO$_2$,\cite{Yang2014a,Ricaurte2014a,Kvamme2007a} as well as to store and transport gases, being especially interesting in the case of hydrogen (H$_2$).~\cite{Mao2002a,Mao2004a,lee2005tuning,veluswamy2014hydrogen,willow2012enhancement,florusse2004stable,strobel2007hydrogen,sugahara2009increasing,davoodabadi2021potential,davoodabadi2021potential,hu2006clathrate} Another relevant aspect in the study of hydrates is the use of additives. The thermodynamic stability conditions of these compounds can be highly modified by the use of additives, also known as hydrate promoters/inhibitors.~\cite{Kang2001a,Delahaye2006a,Anderson2007a,Lee2012a,striolo2019molecular,bui2017evidence,bui2018antiagglomerants,phan2016molecular,naullage2019surfactants,naullage2020slow,jacobson2010amorphous,lederhos1996effective,wu2014effect,wang2021recent,sa2019promoting,torre2012co2} Thermodynamic hydrate promoters are additives that will allow hydrate formation under conditions where it would not normally exist (higher temperatures and lower pressures). Contrarily, the use of inhibitors prevents the formation of these compounds.

The thermodynamics of hydrates has been studied experimentally,~\cite{Sloan2008a,Ripmeester1998a,Ikeda1999a,Udachin2001a,Nakano1998a,Henning2000a,Smith2004a,Kang2008a} theoretically,~\cite{Platteeuw1957a,Platteeuw1959a} and by molecular simulation.~\cite{Miguez2015a,Perez-Rodriguez2017a,Fernandez-Fernandez2019a,Grabowska2022a,Conde2010a,Conde2013a,Michalis2015a,Costandy2015a,Algaba2023b} Thus, the stability conditions of these compounds are well established. Molecular simulation has the advantage of studying these systems from a microscopic perspective, given information about not only the thermodynamics, including phase equilibria, and structure but also the dynamic of the growing/melting of these compounds. However, molecular simulation studies have to be performed carefully due to a series of approximations/limitations that have to be taken into account.~\cite{Frenkel2002a,Allen1987a} One of the most important limitations is the size of the system that can be simulated. Realistic systems, with an Avogadro number order of molecules, are out of the possibilities of the current supercomputers. Typically, the size of the systems studied through molecular dynamic (MD) or Monte Carlo (MC) simulations goes from a few molecules to $\sim$10$^5$. This manuscript is the third part of a series of papers devoted to investigate the effects of finite-size and dispersion interactions on the three-phase equilibria of CH$_{4}$ and CO$_{2}$ hydrates from computer simulation. For a detailed analysis of the finite-size effect on the dissociation temperatures of CO$_{2}$ and CH$_{4}$ hydrates, we recommend the reader the two first papers of the series.~\cite{paperI,paperII}

To keep the system in a condensed phase and prevent the molecules to escape from the simulation box, it is necessary to use a potential that works in the same way as a container (boundary conditions). In order to avoid the so-called border effects, that take place when the molecules ``feel'' the border of the simulation box, periodic boundary conditions (PBC) are used routinely in simulation.~\cite{Frenkel2002a,Allen1987a} This technique implies to replicate the simulation box in the space. The real simulation box is in the middle of a network conformed by identical images of the original simulation box. If one molecule escapes through one side of the box, the same molecule enters from the opposite side of the box, from one of the replicated images. The PBC technique allows to keep constant the number of molecules inside the simulation box avoiding border effects with non-extra computational effort. As a consequence of the infinite number of replicated boxes, two considerations have to be taken into account. The first one is that a molecule is going to interact only with the closest molecule image among all the possible replicated simulation boxes (minima image criteria). The second one is that to avoid the interaction of a molecule with itself in one of the replicated boxes, the interaction range or cut-off distance, $r_{c}$, cannot be larger than half of the shortest side of the simulation box. 

One of the most important parameters needed to evaluate the dispersive interactions between molecules in simulation is the cut-off distance. The election of this cut-off value plays a key role in molecular simulation. Its value governs the range of the interactions between the molecules forming the system under study. Two chemical groups of the same or different molecules will interact if the separation between them is smaller than $r_{c}$. In other words, the interactions between chemical groups separated further than the cut-off distance are neglected. Since its value impacts on the determination of the energy of the system, all the thermodynamic, structural, and dynamical properties could be hardly affected by its particular election.The larger the value of $r_{c}$, the better the accuracy of the calculated properties associated with the full intermolecular potential.  For instance, one of the key properties affected by the election of the cut-off distance is the dissociation temperature of the hydrate, $T_{3}$. Short values provoke less attractive interactions in the system, and consequently, a less stable hydrate. This impact is directly reflected in the $T_{3}$ obtained from computer simulation.

The use of high cut-off values in simulations entails large computational efforts. According to this, it is necessary to find a compromise between accuracy and the CPU time needed in the calculations. Typically, the cut-off distances used in the literature to study hydrates range from $0.9$ to $1.1\,\text{nm}$.~\cite{Miguez2015a,Perez-Rodriguez2017a,Fernandez-Fernandez2019a,Grabowska2022a,Conde2010a,Michalis2015a,Costandy2015a} 
Very recently, some of us have analyzed the effect of $r_{c}$ on the three-phase coexistence temperature, $T_{3}$, at a given pressure of the CO$_{2}$ and CH$_{4}$ hydrates from computer simulation.~\cite{Grabowska2022a,Algaba2023a} As far as we know, these works are the only studies in the literature that consider the effect of the cut-off distance on the $T_3$ of hydrates. Particularly, the dissociation temperatures, at $400\,\text{bar}$, have been determined using $r_{c}=1.0$ and $1.9\,\text{nm}$, for the case of the CO$_{2}$ hydrate, and $r_{c}=0.9$ and $1.7\,\text{nm}$ for the CH$_{4}$ hydrate. The effect of changing the cut-off distance is small but not negligible: $T_{3}$ increases from $290$ to $292\,\text{K}$ in the case of the CO$_{2}$ hydrate, and decreases from $297$ to $295\,\text{K}$ in the case of the CH$_{4}$ hydrate. 


The goal of this work is to understand, from a molecular perspective, the effect of the cut-off distance of the dispersive interactions on the dissociation temperature of hydrates. To this end, we combine molecular dynamics simulations and the well-known direct coexistence technique to determine the $T_{3}$ values, at several pressures, of the CO$_{2}$ and CH$_{4}$ hydrates using different cut-off values. We also perform additional independent simulations to assess the effect of using homogeneous and inhomogeneous long-range corrections on the dispersive interactions on the dissociation temperatures of both hydrates.

The rest of the paper is organized as follows: In Sec. II, we describe the simulation details and the molecular models used in this work. In Sec. III we describe the methodology used to analyze the effect of the cut-off on the \textit{T$_3$} at different pressures of the CO$_2$ and CH$_4$ hydrates. The results obtained, as well as their discussion, are described in Sec. IV. Finally, conclusions are presented in Sec. V

\section{Simulation details and molecular models}

\subsection{Simulation details}

MD simulations are carried out through the GROMACS package.\cite{VanDerSpoel2005a} Here it is important to mention that all simulations have been carried out using GROMACS version 4.6 in double precision except for simulations where the particle mesh Ewalds long-range corrections\cite{Essmann1995a,Lundberg2016a} for the dispersive interactions have been taking into account. GROMACS version 4.6 doesn't support this long-range correction, and for this reason, simulations have been performed using the version 2016 in double precision. In all cases, the simulations are run using the $NPT$ or isothermal-isobaric ensemble, allowing each side of the simulation box to change independently to keep the pressure constant as well as to avoid any stress from the hydrate solid structure. We use the Verlet-leapfrog\cite{Cuendet2007a} algorithm for solving Newton's equations of motion with a time step of 2 fs. In order to keep the temperature and the pressure constant along the simulation, the Nosé-Hoover thermostat\cite{Nose1984a} and the anisotropic Parrinello-Rahman barostat\cite{Parrinello1981a} are used with a time constant of 2 ps. We have checked that fluctuations of the temperature due to the use of the thermostat, with a time step of $2\,\text{ps}$, are always below $0.1\,\text{K}$. This value is much slower than the temperature steps used in the direct coexistence technique, allowing to use the method with confidence. In the case of the Parrinello-Rahman barostat, a compressibility value of $4.5\times10^{-5}\,\text{bar}^{-1}$ is applied in the three directions of the simulation box.

To account for the effect of the range of the dispersive interactions on the three-phase coexistence temperature of the hydrates studied at a given pressure, we use different cut-off distances for the dispersive and coulombic interactions. Particularly, we consider four different cut-off values: $0.9$, $1.1$, $1.3$, and $1.6\,\text{nm}$. In terms of reduced units with respect to the molecular size of the oxygen atom of water molecules (see Table \ref{Model}), the cut-off values range from $2.8$ to $5$, approximately. Besides, two additional sets of simulations have been performed using two different long-range corrections (LRCs) for the dispersive interactions. The first one is the so-called homogeneous LRCs for the dispersive energy and pressure (LRCH),~\cite{Allen1987a,Frenkel2002a} assuming explicitly that the system is homogeneous. Strictly this is not true because the system exhibits three phases in equilibrium. However, this methodology has been applied previously in water systems that exhibit one or more phases obtaining good results.\cite{Grabowska2022a,Grabowska2022b,Blazquez2020a,Montero2023a,Blazquez2023a} The reason is that the uncertainties incorporated by the LRCH, with the assumption that the system is homogeneous, are smaller than the error introduced when simulations are performed using a small cut-off distance.

The second set of LRCs used in this work is the particle mesh Ewald (PME)\cite{Essmann1995a,Lundberg2016a} method for dispersive interactions (LRCI). The PME method has the advantage that the inhomogeneity of the system is explicitly taken into account. Strictly speaking, this is the appropriate set of LRCs that has to be used in systems that exhibit more than one phase in coexistence. However, there exists a limitation for this methodology when using the GROMACS package: it can be only used when unlike dispersive interactions follows the standard geometric Lorentz-Berthelot combining rules. Unfortunately, as it is shown below, this set of LRCs can be only applied in the study of the CH$_{4}$ hydrate.

Finally, it is important to note, that for the case of the coulombic interactions, LRCs have been applied in all cases. It means that the coulombic potential is truncated with the same cut-off distance as the dispersive interactions, and then, the particle mesh Ewald (PME) method\cite{Essmann1995a} is applied to take into account the dismissed coulombic interactions. Hence, the effect of using different cut-off distances for the coulombic interactions is negligible because long-range corrections have been applied in all the simulations. The PME range corrections, for all the interactions, have been used with GROMACS default values (a mesh width of 0.1 nm and a relative tolerance of 10$^{-5}$). 

\subsection{Force Fields}

Water molecules are modeled using the well-known TIP4P/Ice model.~\cite{Abascal2005b} This model has been widely used for the study of the solid phases of water.~\cite{Conde2010a,Miguez2015a,Conde2017a,Grabowska2022a,Grabowska2022b} CO$_2$ molecules are described using the TraPPE force field,\cite{Potoff2001a} and CH$_{4}$ molecules are modeled as single Lennard-Jones (LJ) spheres with molecular parameters taken from the work of Guillot and Guissani.~\cite{Guillot1993a,Paschek2004a} These models have been previously used by some of us for determining the three-phase coexistence lines of CH$_4$ and CO$_2$ hydrates.\cite{Conde2010a,Miguez2015a,Grabowska2022a,Grabowska2022b,paperI,paperII} Non-bonded interactions between different chemical groups are calculated through the sum of the LJ and the Coulomb intermolecular potentials:

\begin{equation}
U(r_{ij})=4\epsilon_{ij}\left[\left(\frac{\sigma_{ij}}{r_{ij}}\right)^{12} -
\left(\frac{\sigma_{ij}}{r_{ij}}\right)^{6}\right] + \frac{q_iq_j}{4\pi\epsilon_0r_{ij}}
\end{equation}

\noindent
where $r_{ij}$ is the distance between the chemical groups $i$ and $j$, $\sigma_{ij}$ and $\epsilon_{ij}$ are the diameter and well depth associated with the LJ potential, $q_i$ and $q_j$ are the partial charges placed on chemical groups $i$ and $j$, and $\epsilon_0$ is the permittivity of vacuum. The molecular parameters associated with each chemical group of water, carbon dioxide, and methane molecules are summarized in Table \ref{Model}. The non-bonded interaction parameters between unlike groups are calculated through the standard Lorentz-Berthelot combining rules except for the water--CO$_2$(2) interactions. In this case, the crossed interaction energy parameter is calculated as $\epsilon_{12}=\xi(\epsilon_{11}\,\epsilon_{22})^{1/2}$, with $\xi=1.13$. This modification of the Berthelot combining rule is necessary in order to accurately predict the three-phase coexistence line of the CO$_2$ hydrate.\cite{Miguez2015a} 

\begin{table}
\caption{Non-bonded interaction parameters of water,~\cite{Abascal2005b} carbon dioxide,\cite{Potoff2001a} and metane.\cite{Guillot1993a,Paschek2004a}}
\centering
\begin{tabular}{lccc}
\hline\hline
Atom &$\sigma(\text{\AA})$ & $\varepsilon/k_B(\text{K})$ & $q$(e) \\
\hline
Water molecule& \\
\hline
O            & 3.1668 & 106.1 & - \\
H & - & - & 0.5897 \\
M  & - & - & 1.1794 \\
\hline
Carbon dioxide molecule& \\
\hline
C & 2.80 & 27.0 & 0.700 \\
O & 3.05 & 79.0 & -0.350 \\
\hline
Methane molecule& \\
\hline
CH$_4$ & 3.73 & 147.5 & - \\
\hline\hline
\end{tabular}
\label{Model}
\end{table}

\section{Methodology}
In this work, we determine the three-phase equilibrium coexistence temperature $T_{3}$, at several pressures, through the direct coexistence simulation technique. This methodology has been widely applied in the literature for determining the stability conditions of hydrates.~\cite{Conde2010a,Miguez2015a,Costandy2015a,Michalis2015a} The technique is based on the simulation of the three phases, i.e., the hydrate, aqueous solution, and guest-rich fluid phases (liquid CO$_2$ and gas CH$_4$ in our particular case), that coexist in the same simulation box and separated by two planar interfaces. According to the phase rule for non-reacting systems, a binary mixture that exhibits three phases in equilibrium has only one thermodynamic degree of freedom. Since the system is inhomogeneous, we fixed the component of the pressure tensor perpendicular to the planar interfaces. Since we also fixed the temperature of the system, there are three possible scenarios depending on the temperature considered. 

\begin{figure*}
\includegraphics[width=\textwidth]{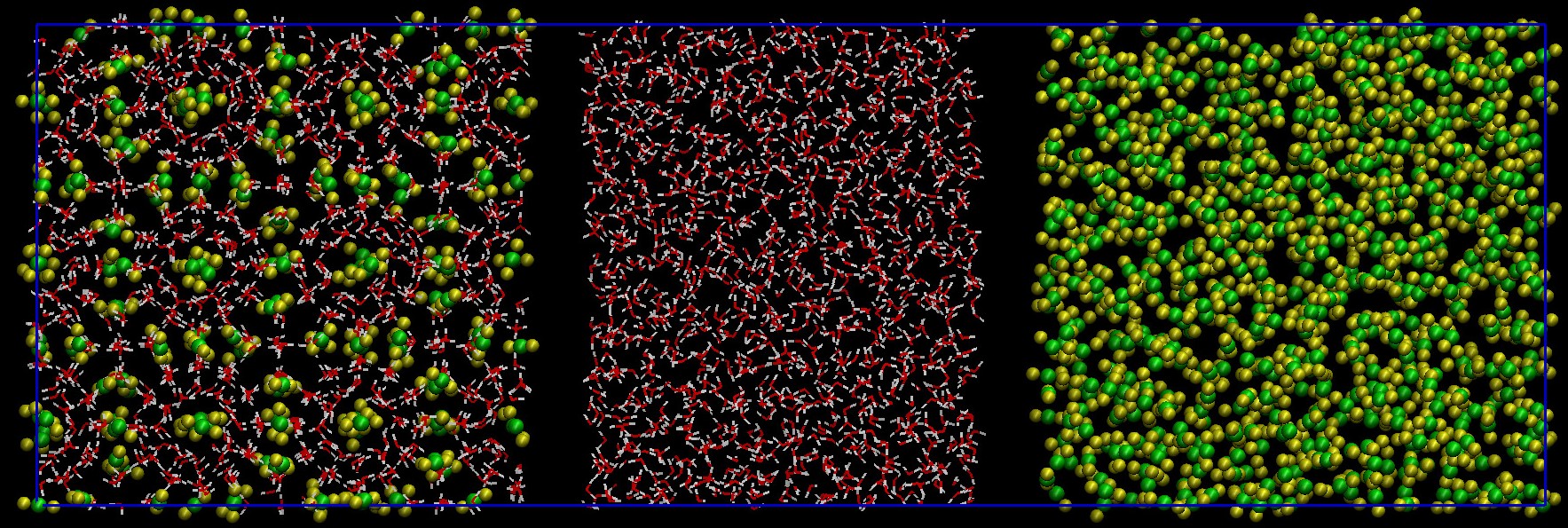}   
\caption{Representation of the initial simulation box used in this work for the case of the CO$_2$ hydrate. Red and white spheres represent the oxygen and hydrogen atoms of the water molecules, and green and yellow spheres carbon and oxygen atoms of the CO$_2$ molecule, respectively. From left to right, the simulation box is conformed by a hydrate phase, a water phase, and a CO$_2$ phase. For the case of the CH$_4$ hydrate, the initial simulation box is identical but the CO$_2$ is replaced by CH$_4$.}
\label{simulationbox}
\end{figure*}

If the temperature is above the $T_3$, the hydrate will melt and only two phases remain stable in the simulation box: the aqueous solution and the CO$_{2}$- or the CH$_{4}$-rich fluid phase. Contrary, if the temperature is below the $T_3$, the hydrate will grow until one of the fluid phases disappears (the aqueous solution or the guest-rich fluid phase). Which fluid phase is the remaining depends on the amount of each component used in the simulation box. Finally, there is a third scenario when the temperature of the system matches the $T_3$. When the system is simulated at the three-phase equilibrium conditions, the hydrate will grow/melt with a 50\% probability. It means that if more than one simulation is performed at these conditions, the hydrate will grow/melt in half of them. Since we only need to know if the simulated temperature is above (the hydrate melts) or below (the hydrate grows) the dissociation temperature of the hydrate, $T_{3}$, we are not interested in the equilibrium state of the system. This allows to locate, within the corresponding error bars, the $T_{3}$ of the hydrate.

Although this procedure has been used in previous works for the determination of the three-phase coexistence line of the CH$_{4}$ and CO$_{2}$ hydrates, the effect of the range of the dispersive interactions on the $T_3$ has not been properly analyzed. Usually, the size of the hydrate used in the initial simulation boxes in previous works has been $2\times 2\times 2$,~\cite{Conde2010a,Miguez2015a,Costandy2015a,Michalis2015a} which means that the hydrate unit cell has been replicated 2 times in each space direction (368 and 64 molecules of water and guest respectively).

In this work, we use an initial simulation box in which a hydrate phase fully occupied is in contact with a pure water phase and a phase formed from CO$_{2}$ or CH$_{4}$ molecules depending on the hydrate considered. The hydrate phase is created by replicating 3 times the unit cell in each space direction. The final $3\times 3\times 3$ hydrate phase is formed by $1242$ and $216$ water and guest molecules, respectively. The initial water phase is formed by $1242$ water molecules, and the guest phase by $400$ molecules (see Table \ref{tabla-moleculas}). Note that the use of this initial simulation box pursues two objectives: (1) to avoid finite-size effects; and (2) to allow the use of cut-off distances up to $1.6\,\text{nm}$.  A sketch of the initial simulation box used in this work is presented in Fig.~\ref{simulationbox} It is important to notice that the election of the final system size is far from being arbitrary. According to the two first papers of the series,~\cite{paperI,paperII} a non-stoichiometric number of water and guest molecules in the initial aqueous solution and guest-rich phases in combination with a $3\times 3\times 3$ hydrate phase allow us to determine accurately the $T_3$ value without finite-size effects (we refer the lector to papers I\cite{paperI} and II\cite{paperII} for further details). Also, it is interesting to mention that the empty space between phases shown in Fig. \ref{simulationbox} is rapidly occupied by the water solution and guest-rich phases in the first steps of the simulation. The initial gap between phases prevents the overlapping of the molecules of each phase when they are placed together in the same simulation box.

\begin{table}
\begin{tabular}{c c c c c c c c c c c c c c c}
\hline
\hline
 &   \multicolumn{5}{c}{Hydrate phase} & &
{Water solution} & &
{Guest-rich phase} \\
     \cline{2-6}
     & Unit Cell &  & Water &  & CO$_2$/CH$_4$ &  & Water &  & CO$_2$/CH$_4$ \\
\hline
& 3x3x3 & & 1242 & &  216  & & 1242  & & 400 \\
\hline
\hline
\end{tabular}
\caption{\label{tabla-moleculas} Initial number of molecules in the hydrate phase, water solution phase, and guest-rich phase. The same configuration for methane and carbon dioxide has been used in all the simulations carried out in this work. } 
\end{table}

\section{Results}

We have performed simulations at three different pressures in the case of the CO$_2$ hydrate: $100$, $400$, and $1000\,\text{bar}$. However, we only perform simulations at $400\,\text{bar}$ in the case of the CH$_{4}$ hydrate. The reason is that simulation times needed to observe growth or melt in the case of the CH$_{4}$ hydrate are much longer than in the case of the CO$_{2}$ hydrate due to the low solubility exhibited by CH$_{4}$ in water. In both cases, the $T_3$ has been obtained using 4 different cut-off distances for the dispersive interactions: $0.9$, $1.1$, $1.3$, and $1.6\,\text{nm}$.  The same cut-off distance is used for the coulombic part of the intermolecular potential. The cut-off distance values used in this work have been carefully selected. The two lowest, $0.9$ and $1.1\,\text{nm}$ have been previously used by several authors to determine dissociation temperatures of CH$_{4}$ and CO$_{2}$ hydrates.~\cite{Conde2010a,Conde2013a,Michalis2015a,Miguez2015a,Costandy2015a} The $1.6\,\text{nm}$ value has been also used by some of us to calculate the dissociation temperature of the same hydrates.~\cite{Grabowska2022a,Algaba2023a} Finally, $1.3\,\text{nm}$ is an intermediate value that helps to get a general picture of how $T_{3}$ varies as the cut-off distance values increase. As it has been explained in Section II, we apply PME corrections for the coulombic interactions. We also analyze the effect of using LRCs on the dissociation temperature of each hydrate. To this end, we have carried out additional simulations using two different schemes to account for the LRCs of the dispersive interactions with a cut-off distance equal to $0.9\,\text{nm}$. Particularly, we use the standard LRCH corrections for both hydrates, and in the case of the CH$_{4}$ hydrate, we also used the LRCI corrections.

\begin{figure*}
     \centering
         \centering
         \includegraphics[width=0.30\textwidth]{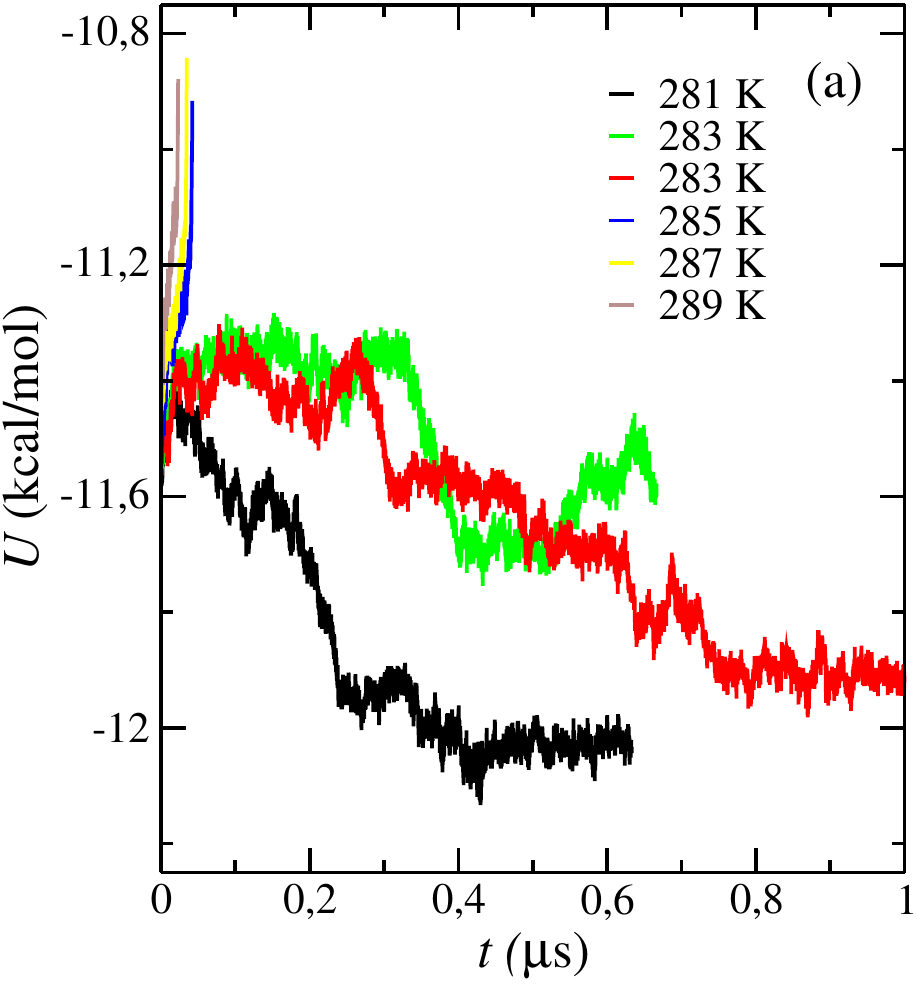}
         \includegraphics[width=0.30\textwidth]{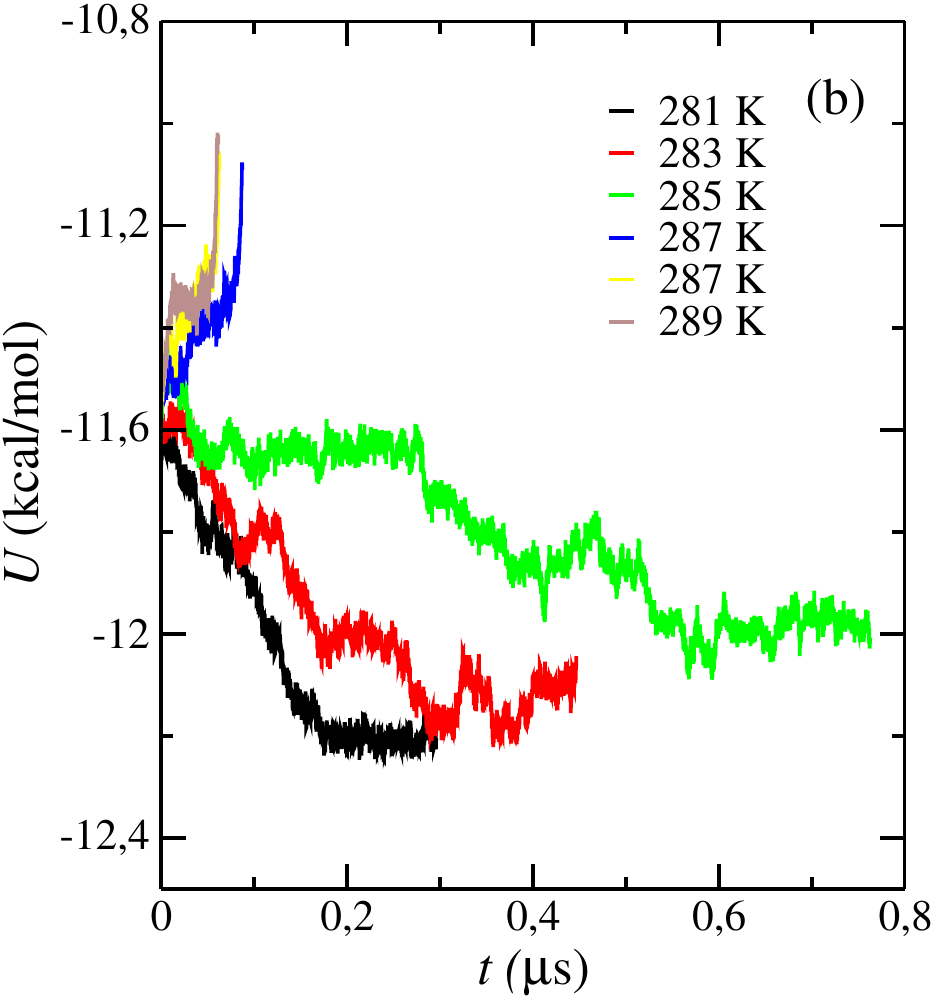}
         \includegraphics[width=0.30\textwidth]{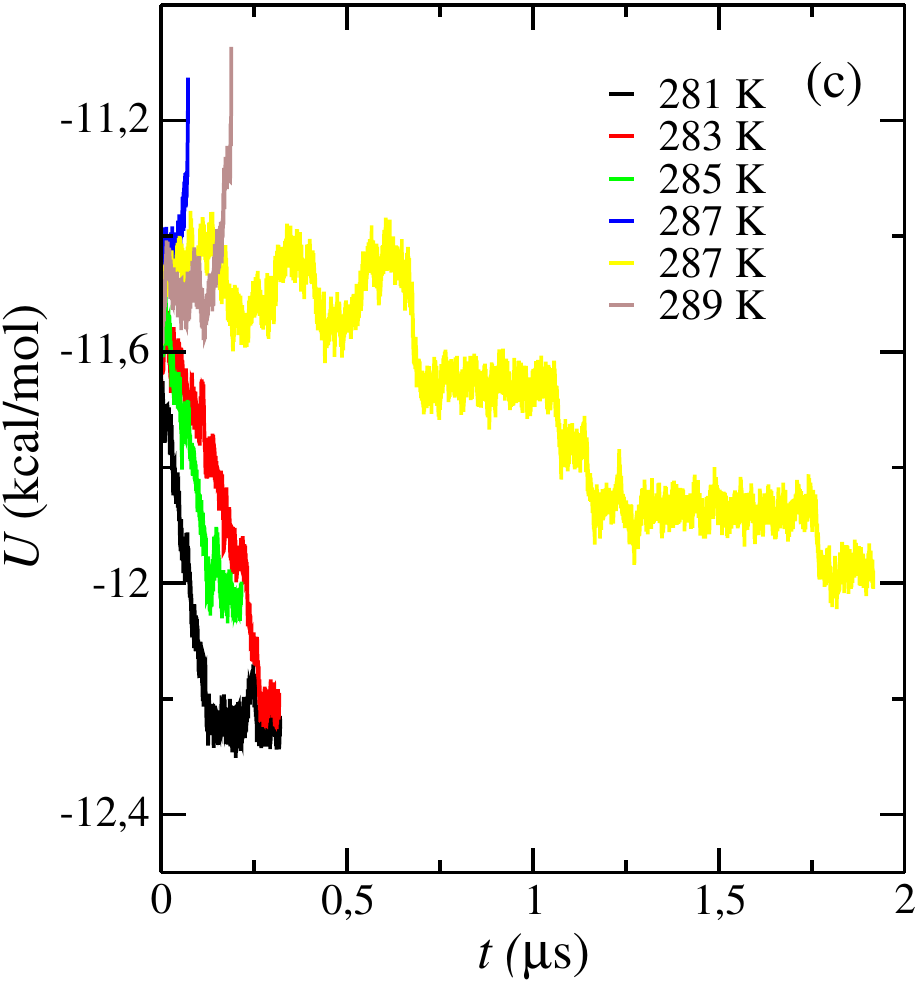}
         \includegraphics[width=0.30\textwidth]{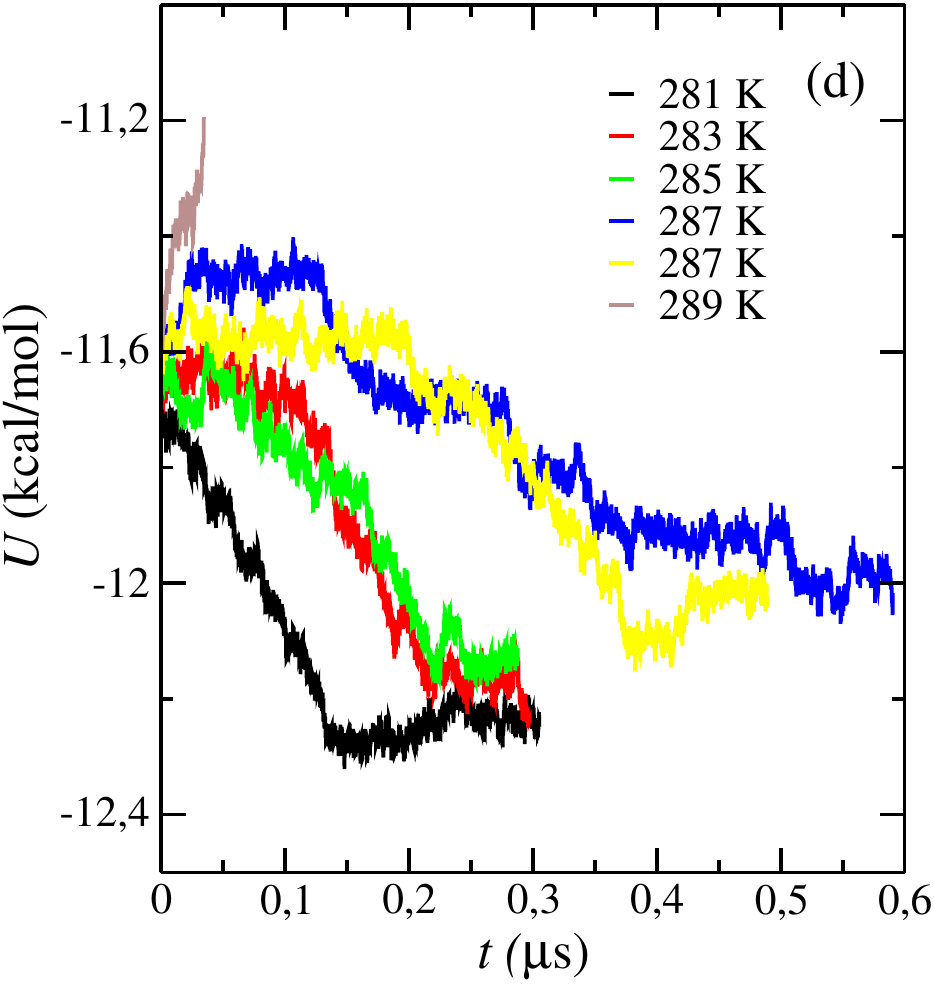}
         \includegraphics[width=0.30\textwidth]{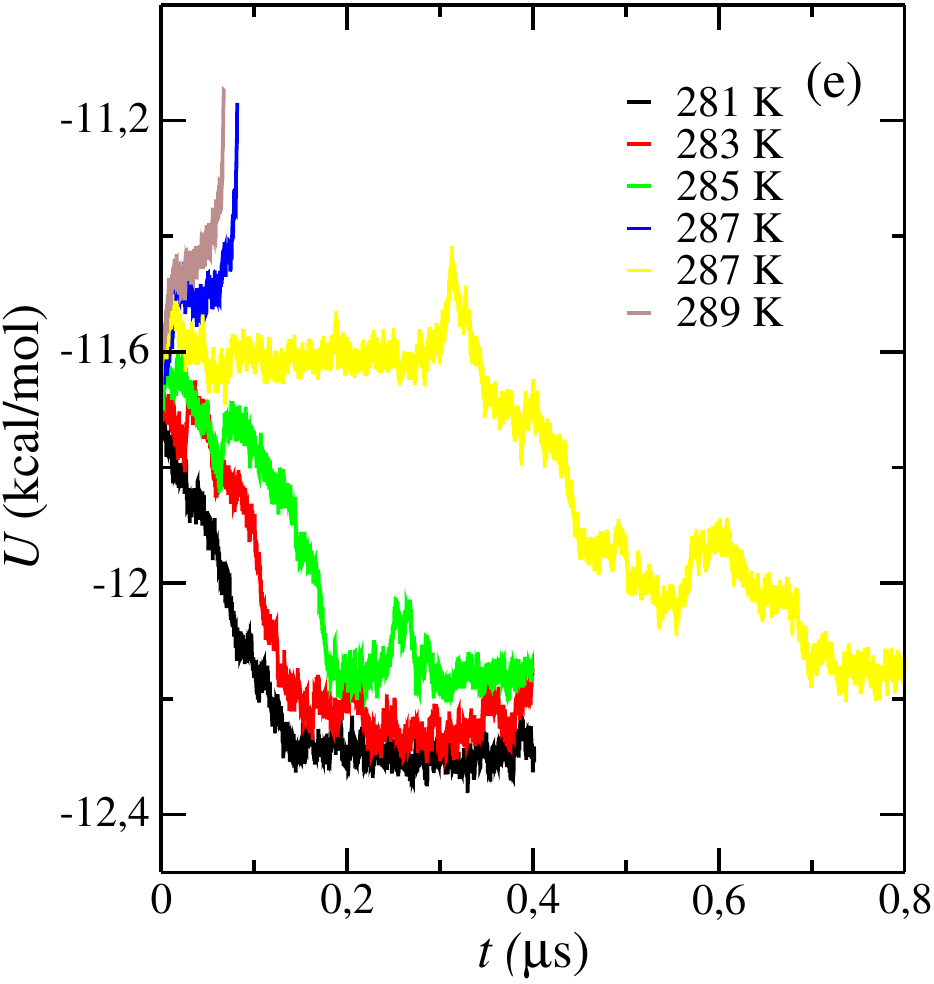}
\caption{Potential energy, as a function of time, at $100\,\text{bar}$ and different temperatures for different cut-off values: (a) $0.9\,\text{nm}$, (b) $1.1\,\text{nm}$, (c) $1.3\,\text{nm}$, (d) $1.6\,\text{nm}$, and (e) $0.9\,\text{nm}$ + LRCH. Notice that large simulation times are necessary when the temperature is close to the $T_3$ value.}
\label{CO2_100_simulations}
\end{figure*}

\subsection{CO$_2$ Hydrate}

We first analyze the effect of cut-off distance on the dissociation temperatures of the CO$_{2}$ hydrate. At each pressure, we perform simulations at five temperatures in order to analyze the behavior of the system. Particularly, the temperatures considered are separated by $2\,\text{K}$.  Due to the inherent stochasticity of the direct coexistence technique, at temperatures close to the $T_{3}$ we run two independent simulations for each cut-off value. This is done to increase the accuracy of the determination of the $T_3$.

\begin{figure}
\includegraphics[width=\columnwidth]{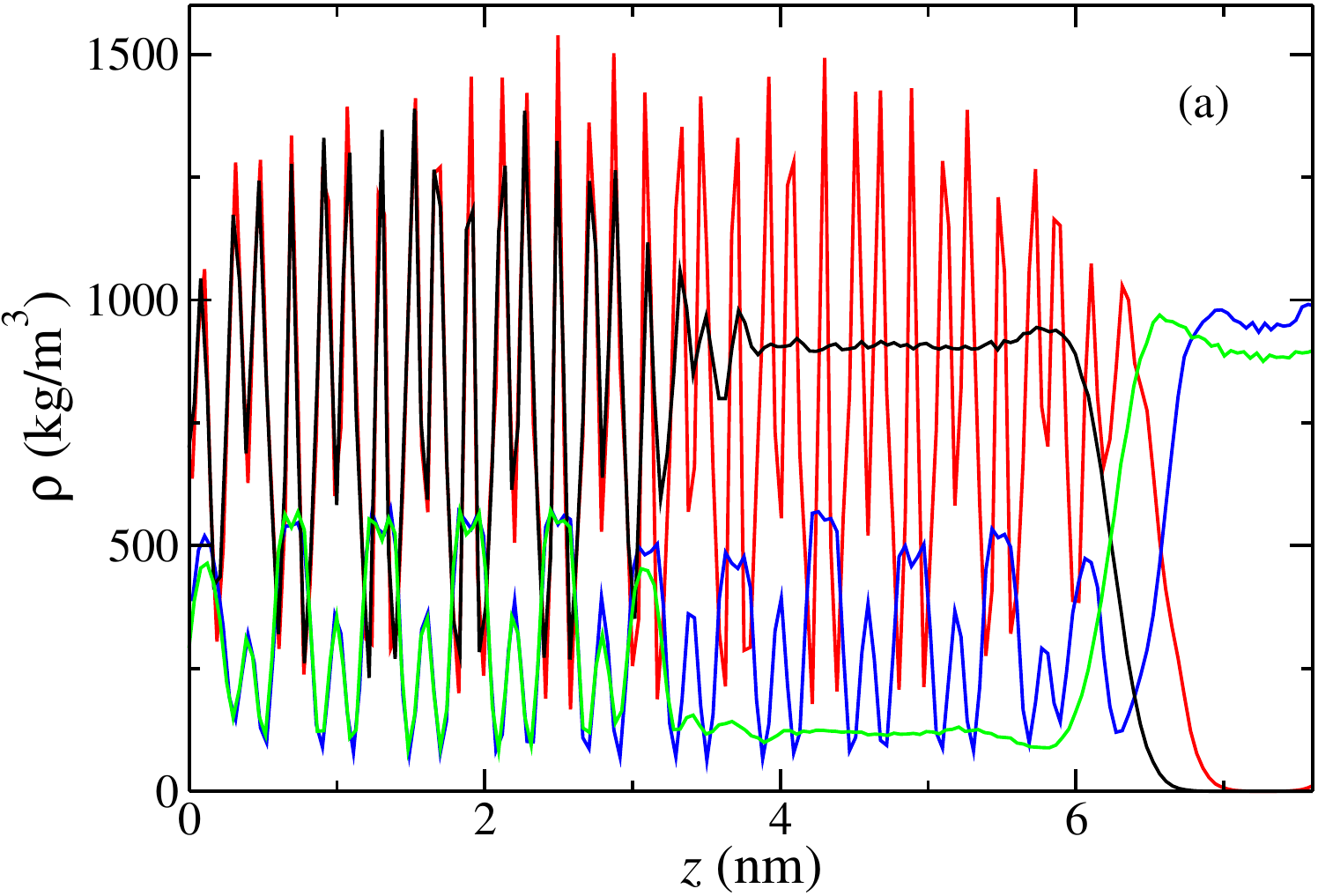}\\
\includegraphics[width=\columnwidth]{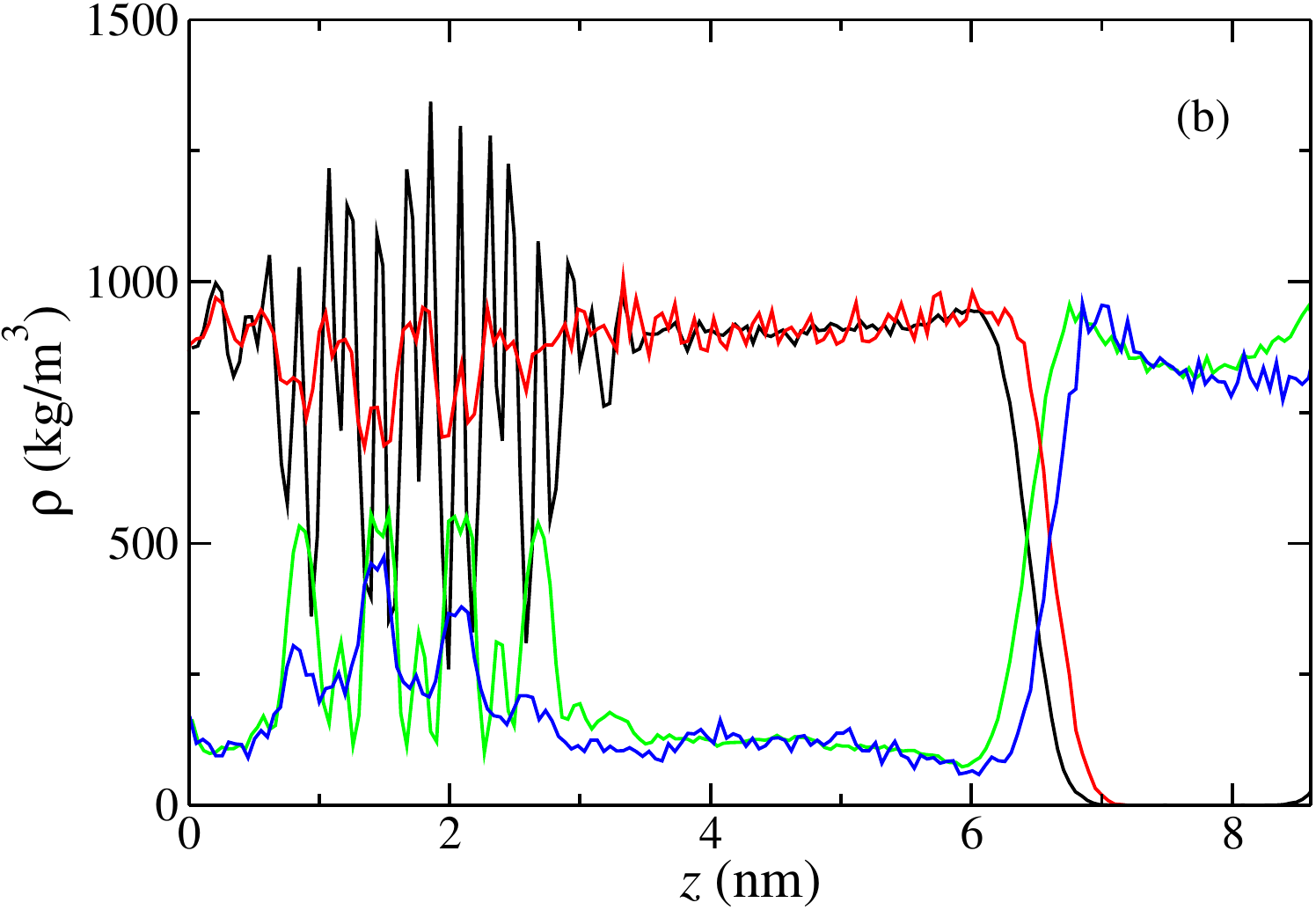}\\

\caption{Density profiles, as obtained from simulations, at $100\,\text{bar}$ and at (a) $281$ and (b) $287\,\text{K}$, using a $r_{c}=1.1\,\text{nm}$. In both cases, black and green lines correspond to the density of water and CO$_2$, respectively at the beginning of the simulations, while red and blue lines correspond to the density of water and CO$_2$, respectively at the end of the simulations.}
\label{density}
\end{figure}



\begin{figure}
\includegraphics[width=\columnwidth]{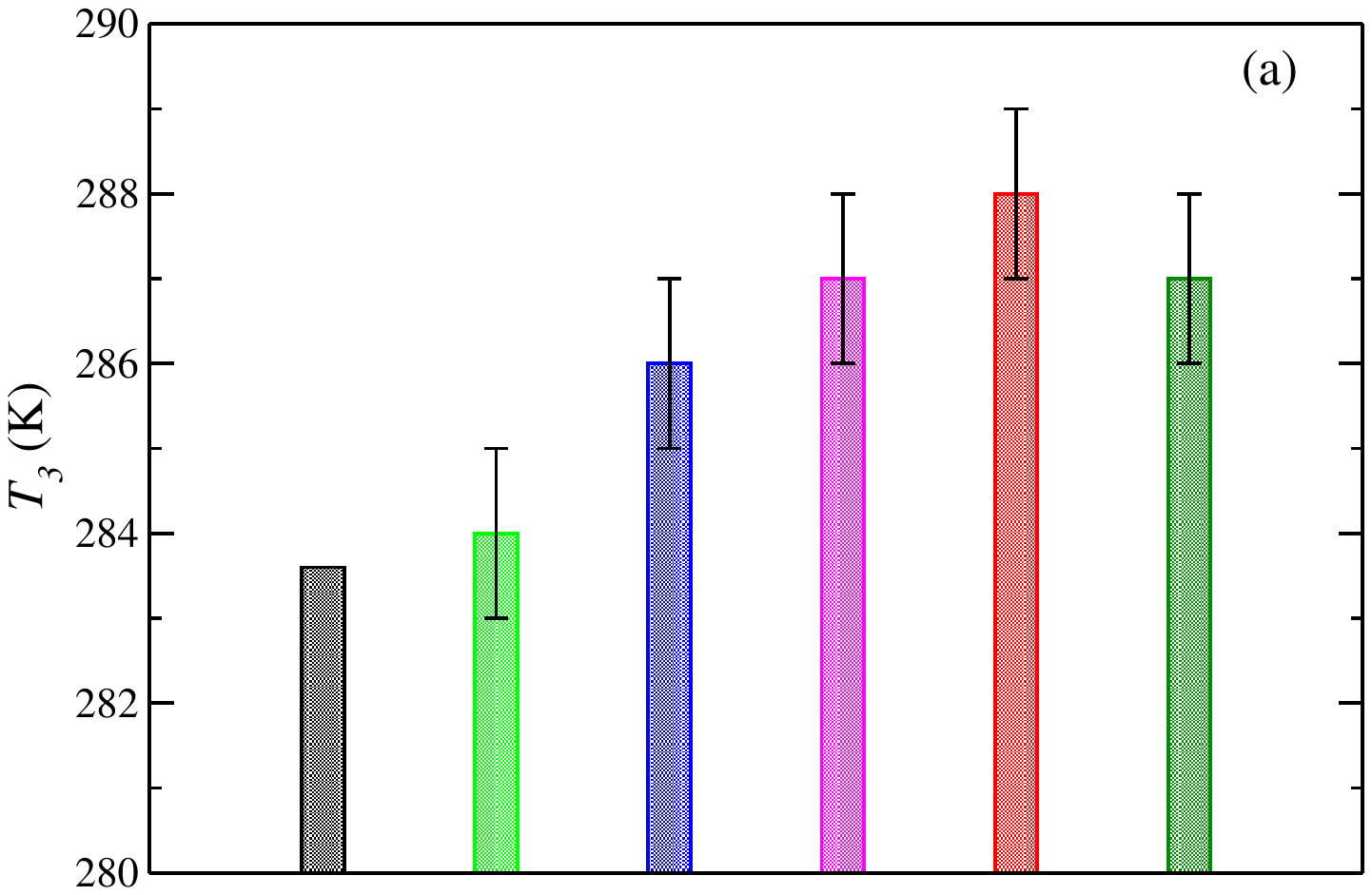}\\
\vspace{-0.6cm}

\includegraphics[width=\columnwidth]{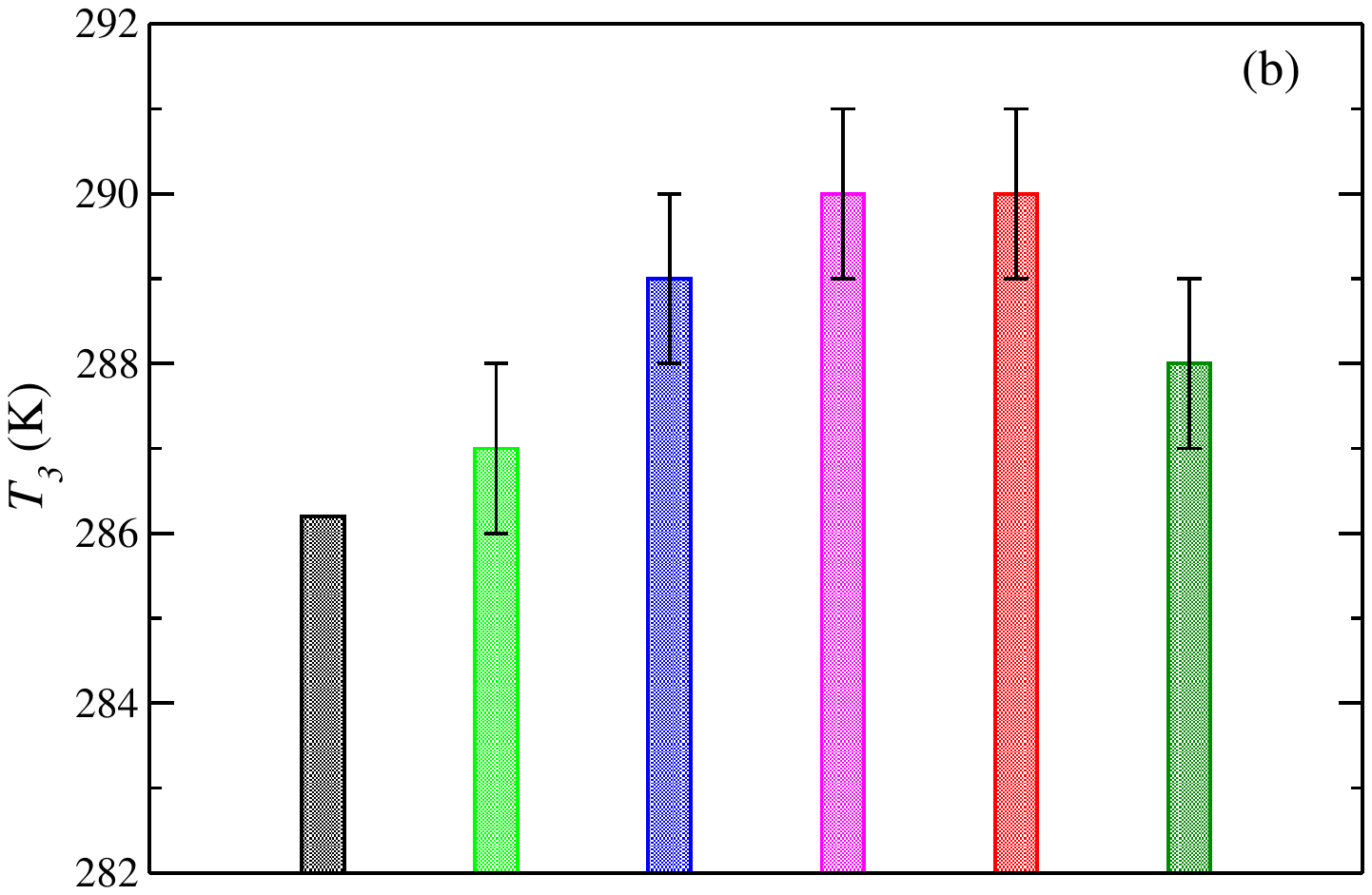}\\
\vspace{-0.6cm}
\includegraphics[width=\columnwidth]{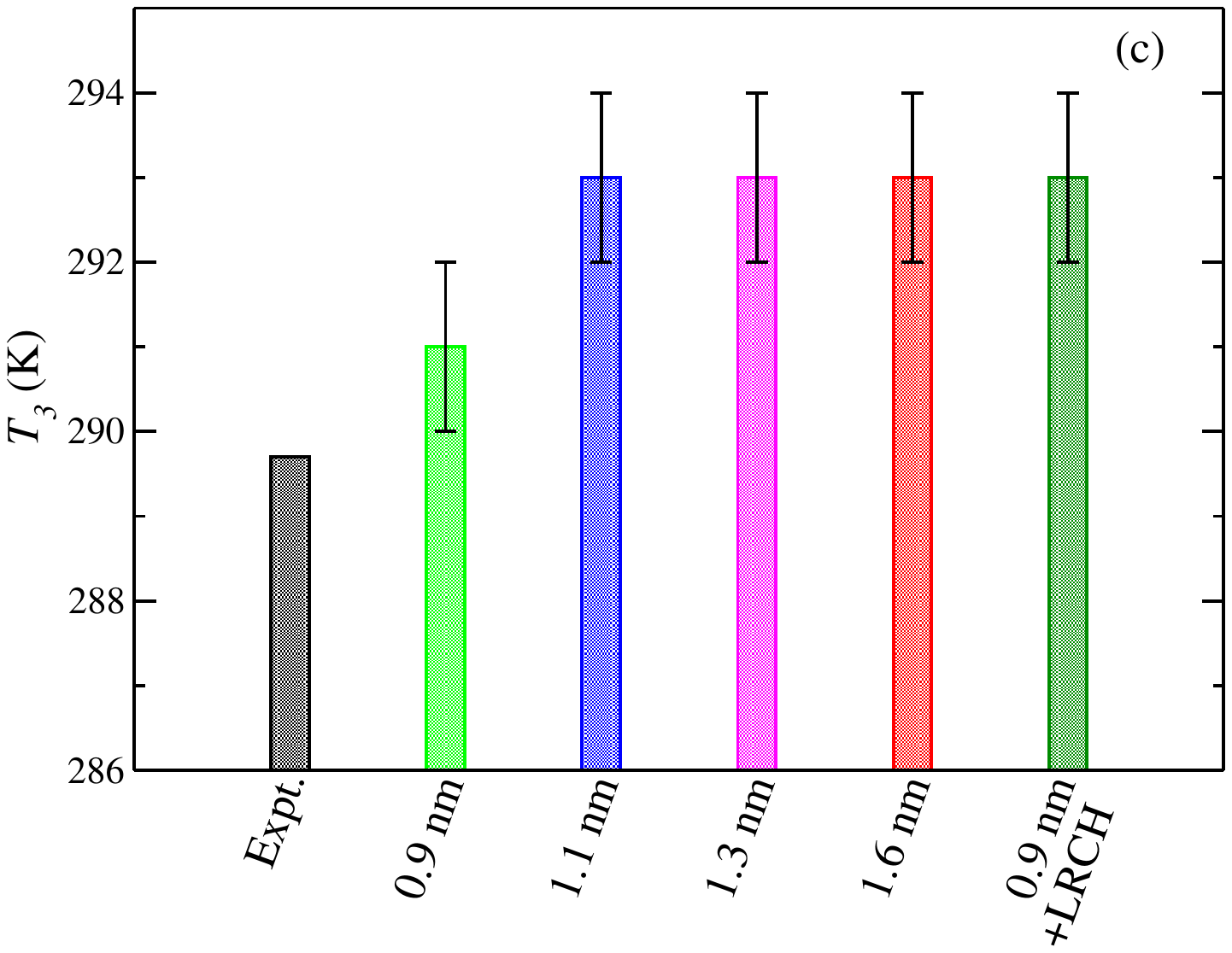}

\caption{Dissociation temperature of the CO$_2$ hydrate, as a function of the cut-off distance, at (a) $100\,\text{bar}$, (b) $400\,\text{bar}$, and (c) $1000\,\text{bar}$. Results obtained in this work are presented as black bars and uncertainties as red lines.}
\label{CO2_result}
\end{figure}

At 100 bar, simulations are performed at temperatures from $281$ to $289\,\text{K}$. As can be seen in Figs.~\ref{CO2_100_simulations} and \ref{density}, the $T_3$ value increases as the cut-off distance is increased. In particular, we obtain $284(1)$, $286(1)$, $287(1)$, and $288(1)$ for cut-off distances of $0.9$, $1.1$, $1.3$, and $1.6\,\text{nm}$, respectively. The main result is that an increase in the range of the dispersive interactions provokes an increase in the stability of the solid phase, allowing the hydrate to be stable at higher temperatures. Particularly, the $T_{3}$ value at $100\,\text{bar}$ increases from $284(1)\,\text{K}$, when a cut-off distance of 0.9 nm is used, to $288(1)\,\text{K}$ when the cut-off distance is $1.6\,\text{nm}$, as it is clearly shown in Fig.~\ref{CO2_result}.

It is possible to provide additional evidences that corroborate the results obtained. We have determined the density profiles of water and CO$_{2}$ molecules along the $z$-axis, perpendicular to the interfaces, at the beginning of the simulation and at times at which the evolution of the potential energy curves indicates melting/growing of the hydrate. Fig.~\ref{density} shows the corresponding density profiles of water and CO$_{2}$ at $100\,\text{bar}$. As can be seen, at $281\,\text{K}$ the hydrate grows and water and CO$_2$ molecules occupy the simulation box in the characteristic sI crystalline solid structure. This is in agreement with the evolution of the potential curve shown in panel b of Fig.~\ref{CO2_100_simulations} (black curve). Contrary, at $287\,\text{K}$ the hydrate melts and water and CO$_2$ molecules occupy the simulation box in two liquid phases, water- and a CO$_{2}$-rich liquid phases. This is in agreement with the evolution of the potential curve shown in panel b of Fig.~\ref{CO2_100_simulations} (blue and pink curves).

It is important to compare the results obtained in this work with the data presented by M\'{\i}guez \emph{et al.}~\cite{Miguez2015a} several years ago, as well as with experimental data. Note that the main goal of this work is not to determine the most suitable cut-off distance to reproduce the experimental dissociation temperature. The use of larger cut-off distances provides better descriptions of the full potential model. A more interesting issue is to know which is the lowest value of $r_{c}$ able to capture the full potential (optimal value of $r_{c}$). Obviously, larger cut-off distance values than this optimal value would provide the same $T_{3}$ values. However, it is also important to know how the $T_{3}$ predicted using larger values of $r_{c}$ compares with experimental data.
The $T_{3}$ value obtained with a cut-off distance equal to $0.9\,\text{nm}$, $284(1)\,\text{K}$, is in excellent agreement with the result obtained by Míguez~\emph{et al.}, $284(2)\,\text{K}$, using a cut-off value of $1.0\,\text{nm}$. The result is also in good agreement with experimental data taken from the literature ($283.6\,\text{K}$).\cite{Sloan2008a} Note that when the cut-off value is increased, the simulation results slightly overestimate the experimental value. Why does agreement between experimental data and simulation results become worse as the cut-off distance increases? The reason is related to the election of the modified Berthelot combining rule. The value of the water-CO$_2$ unlike the dispersion interaction used in this work, $\xi=1.13$, was initially proposed by M\'{\i}guez and collaborators several years ago.~\cite{Miguez2015a} In that work, the cut-off distance for the dispersive interactions used was $r_{c}=1.0\,\text{nm}$. In addition to that, the system size used was much smaller than that used in this work. Note that, not only the cut-off distance affects the location of the dissociation temperature at a given pressure but also the size of the simulated system. We recommend the reader the papers I and II of this series to understand the finite-size effects on phase equilibria of hydrates.~\cite{paperI,paperII}

The effect of using LRCH corrections on the location of the T$_{3}$ has been also analyzed in this work. As explained in Sec.~II A, the use of LRCH is not strictly correct when several phases coexist in the same simulation box, i.e., when the system is inhomogeneous. However, this methodology has been applied before in the literature to describe multi-phase systems improving the accuracy of the results.~\cite{Grabowska2022a,Grabowska2022b,Blazquez2020a,Montero2023a,Blazquez2023a}  This is because the difference between the simulation predictions using the full and a truncated intermolecular potential using a small cut-off value is bigger than using the LRCH corrections assuming a homogeneous system. As can be seen in Fig.~\ref{CO2_result}, the use of the LRCH with a cut-off value of $0.9\,\text{nm}$ increase significantly the prediction of the $T_3$, from $284(1)\,\text{K}$ ($r_{c}=0.9\,\text{nm}$ and no corrections), to $287(1)\,\text{K}$ ($r_{c}=0.9\,\text{nm}$ and LRCH corrections). Note that the value of $T_3$ obtained is the same, within the error bars as those obtained using $r_{c}=1.3$ and $1.6\,\text{nm}$.

Before finishing the analysis of the results obtained at $100\,\text{bar}$, it is important at this point to remark that an increase in the cut-off value implies an increment of the computational effort necessary to perform the simulations. Indeed, LRCs are routinely used to reduce the computational effort without lost accuracy. Particularly, simulations performed using $r_{c}=0.9\,\text{nm}$ with and without LRCH require the same CPU time. However,
simulations performed using a cut-off value of $0.9\,\text{nm}$ are, on average, three times faster than simulations with a cut-off value of 1.6 nm.

Following the same procedure used to determine the dissociation temperature of the CO$_{2}$ hydrate at $100\,\text{bar}$, we perform simulations at temperatures from $284$ to $292\,\text{K}$ to estimate the $T_{3}$ at $400\,\text{bar}$ using different cut-off distances. Particularly, we obtain $287(1)$, $289(1)$, $290(1)$, and $290(1)\,\text{K}$ for cut-off distances equal to $0.9$, $1.1$, $1.3$, and $1.6\,\text{nm}$, respectively. As can be seen, there is no influence of the cut-off value on the results when using $r_{c}=1.3$ and $1.6\,\text{nm}$. It is also interesting to note that the difference between the results obtained using $0.9$ and $1.6\,\text{nm}$ cut-off distances is smaller now ($3$ instead of $4\,\text{K}$ obtained at $100\,\text{bar}$). Although the $T_{3}$ values are within the uncertainties bars, it seems that the cut-off value used to deal with the dispersive interactions has less effect on $T_{3}$ as the pressure increases. It is also important to compare the results obtained in this work with the predictions previously presented in the literature, as well as with experimental data taken from the literature. We recall again that we are not determining the most suitable cut-off distance to reproduce the experimental dissociation temperature. The $T_{3}$ value obtained with a cut-off distance of $0.9\,\text{nm}$, $287(1)\,\text{K}$, is in excellent agreement with the result obtained by Míguez~\emph{et al.},~\cite{Miguez2015a} $287(2)\,\text{K}$. The $T_{3}$ of the CO$_2$ hydrate at 400 bar has been also obtained very recently by some of us using the so-called solubility method.~\cite{Algaba2023a} In this work, the $T_{3}$ values obtained were $290(2)$ and $292(2)\,\text{K}$ for cut-off distances equal to $1.0$ and $1.9\,\text{nm}$, respectively. Although all the results are in good agreement, within the uncertainties bars, care must be taken when comparing with data from the literature obtained using different system sizes. As we have already discussed in our previous papers I and II, the finite-size effects can affect the value of the dissociation temperature of hydrates obtained from computer simulation.~\cite{paperI,paperII}

It is also interesting to compare the values of the $T_{3}$ obtained using different cut-off distances with experimental data taken from the literature. At  $400\,\text{bar}$, $T_{3}=286.2\,\text{K}$.~\cite{Sloan2008a} As it happens at $100\,\text{bar}$, when the cut-off value is increased the simulation results slightly overestimate the experimental data. As we have mentioned previously, this is due because the Berthelot rule was adjusted using a cut-off distance $r_{c}=1.0\,\text{nm}$\cite{Miguez2015a} and we now are using higher cut-off distances with the same $\xi$ parameter value.

We have also analyzed the impact of using LRCH corrections on the determination of the dissociation temperature of the hydrate at 400 bar. As it is shown in Fig.~\ref{CO2_result}, the dissociation temperature obtained using a cut-off distance of $0.9\,\text{nm}$ and LRCH is $288(1)\,\text{K}$. This value is slightly outside the error bar when it is compared with the data obtained using $r_{c}=1.6\,\text{nm}$.

To finish this Section, we finally consider the effect of the cut-off distance on the dissociation temperature of the hydrate at $1000\,\text{bar}$. To this end, we have performed simulations at temperatures from $286$ to $294\,\text{K}$. At this pressure, the effect of the cut-off distance seems to be negligible for values $r_{c}\ge 1.1\,\text{nm}$. Particularly, $T_{3}=291(1)\,\text{K}$ when using $r_{c}=0.9\,\text{nm}$ and $T_{3}=293(1)\,\text{K}$ when using all of the rest cutoff distances, $r_{c}=1.1$, $1.3$, and $1.6\,\text{nm}$. The dissociation temperature of the hydrate is also equal to $293(1)\,\text{K}$ when the cut-off distance is equal to $0.9\,\text{nm}$ using LRCH. All the results obtained at $1000\,\text{bar}$, are summarized is Fig.~\ref{CO2_result}. It is also interesting to compare the predictions obtained in this work with the data calculated by some of us in a previous work,~\cite{Miguez2015a} as well as with experimental data taken from the literature.~\cite{Sloan2008a} The dissociation temperature of the CO$_{2}$ hydrate at $1000\,\text{bar}$ obtained by M\'{\i}guez \emph{et al.}~\cite{Miguez2015a} is $T_{3}=289(2)\,\text{K}$ and the experimental temperature at the same condition $T_{exp}=289\,\text{K}$. As can be seen, the results obtained in this work slightly overestimate the previous results, as well as the experimental value. This disagreement between the data obtained in this work and previous results of M\'{\i}guez \emph{et al.}~\cite{Miguez2015a} and experimental data~\cite{Sloan2008a} are directly related to the effect of dispersive interactions of the intermolecular potential used in the simulations. The $T_3$ values obtained in this work for the CO$_2$ hydrate at each pressure and each cut-off have been summarized in Table \ref{tabla-T3}.

\begin{table}
\footnotesize
\begin{tabular}{c c c c c c c c c c c c c c c c}
\hline
\hline
\multicolumn{16}{c}{CO$_2$ Hydrate $T_3$ (K)} \\
\hline
$P$ (bar) & & Expt. & & 0.9 nm & & 1.1 nm & & 1.3 nm & & 1.6 nm & & 0.9 nm & & 0.9 nm &\\
\multicolumn{12}{c}{}& +LRCH & & +LRCI &\\
\hline
100 & &283.6 & &284(2) & &286(2) & &287(2) & &288(2) & &287(2) & &- &\\
400 & &286.2 & &287(2) & &289(2) & &290(2) & &290(2) & &288(2) & &- &\\
1000 & &289.7 & &291(2) & &293(2) & &293(2) & &293(2) & &293(2) & &- & \\
\hline
\multicolumn{16}{c}{CH$_4$ Hydrate $T_3$ (K)} \\
\hline
400 & &297 &  &291(1) & &291(1) & &293(1) & &295(1) & &294(1) & &294(1) &\\
\hline
\hline
\end{tabular}
\caption{\label{tabla-T3} Dissociation temperature of the CO$_2$ and CH$_4$ hydrates as a function of the cut-off value at each pressure studied in this work.} 
\end{table}

\begin{figure*}[hbt!]
     \centering
         \centering
         \includegraphics[width=0.30\textwidth]{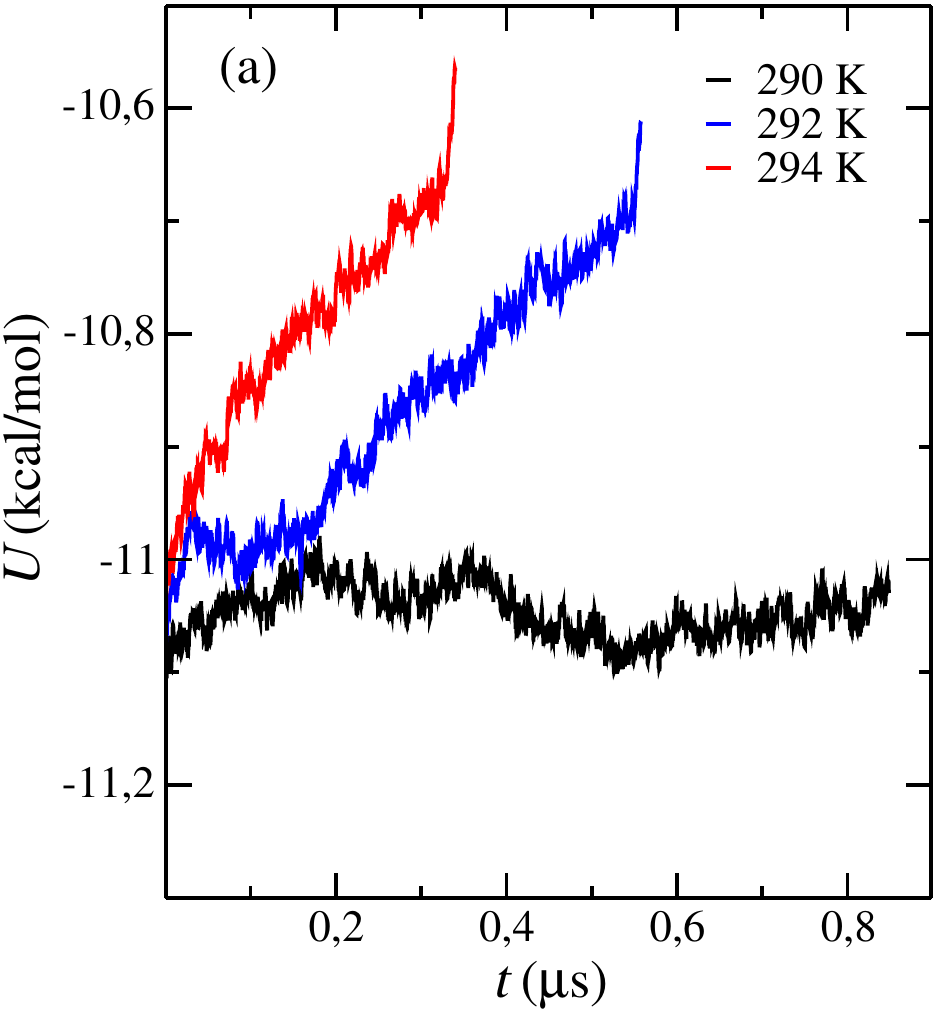}
         \includegraphics[width=0.30\textwidth]{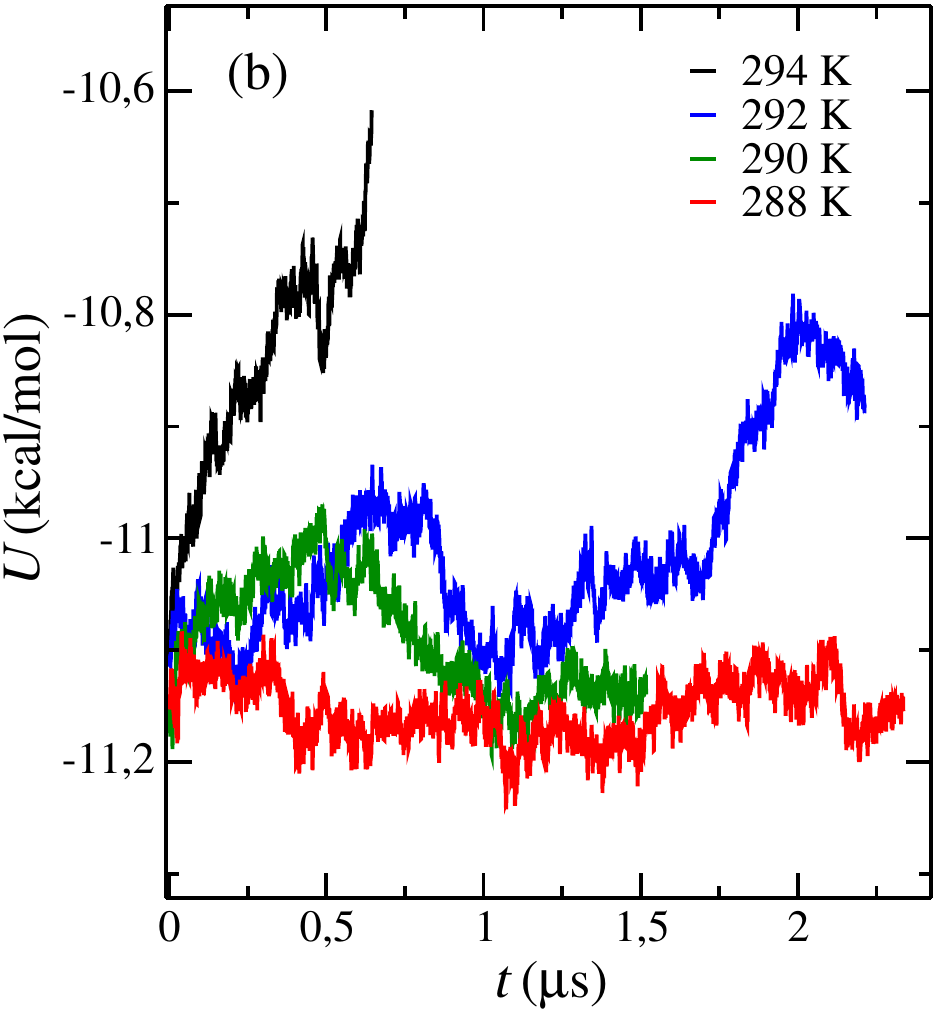}
         \includegraphics[width=0.30\textwidth]{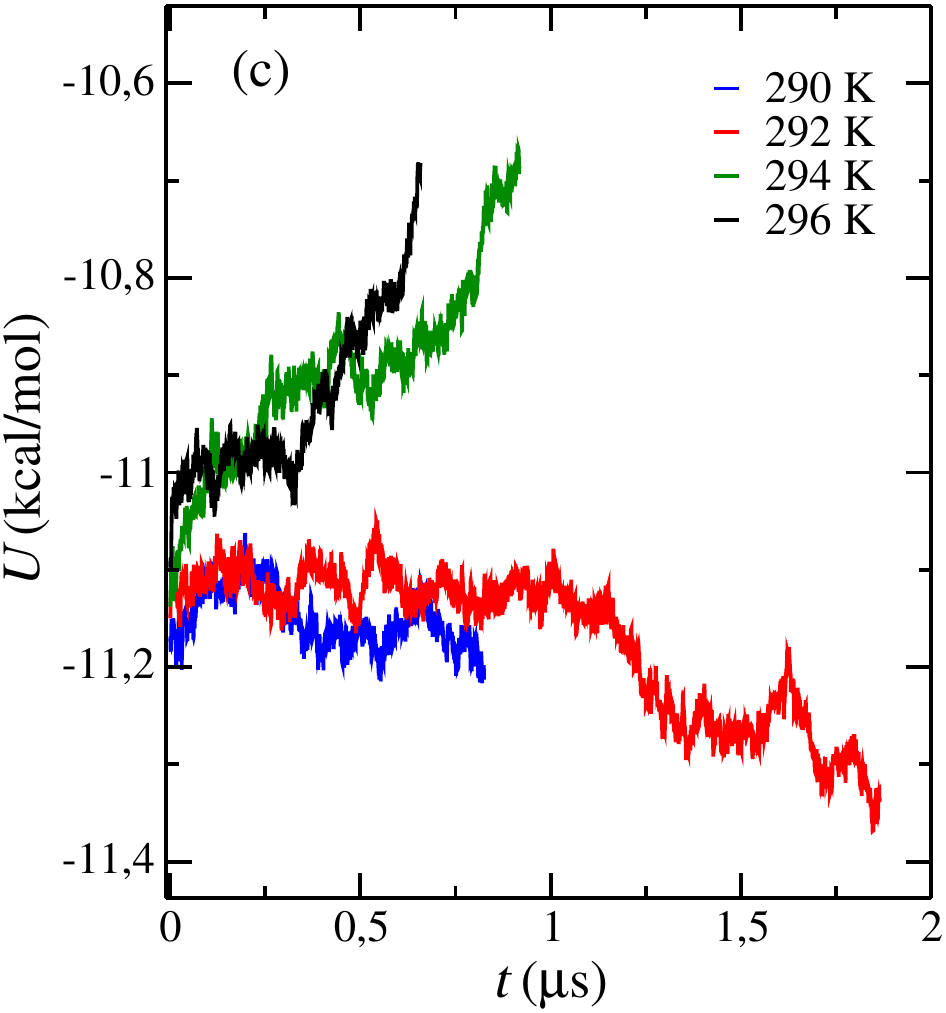}
         \includegraphics[width=0.30\textwidth]{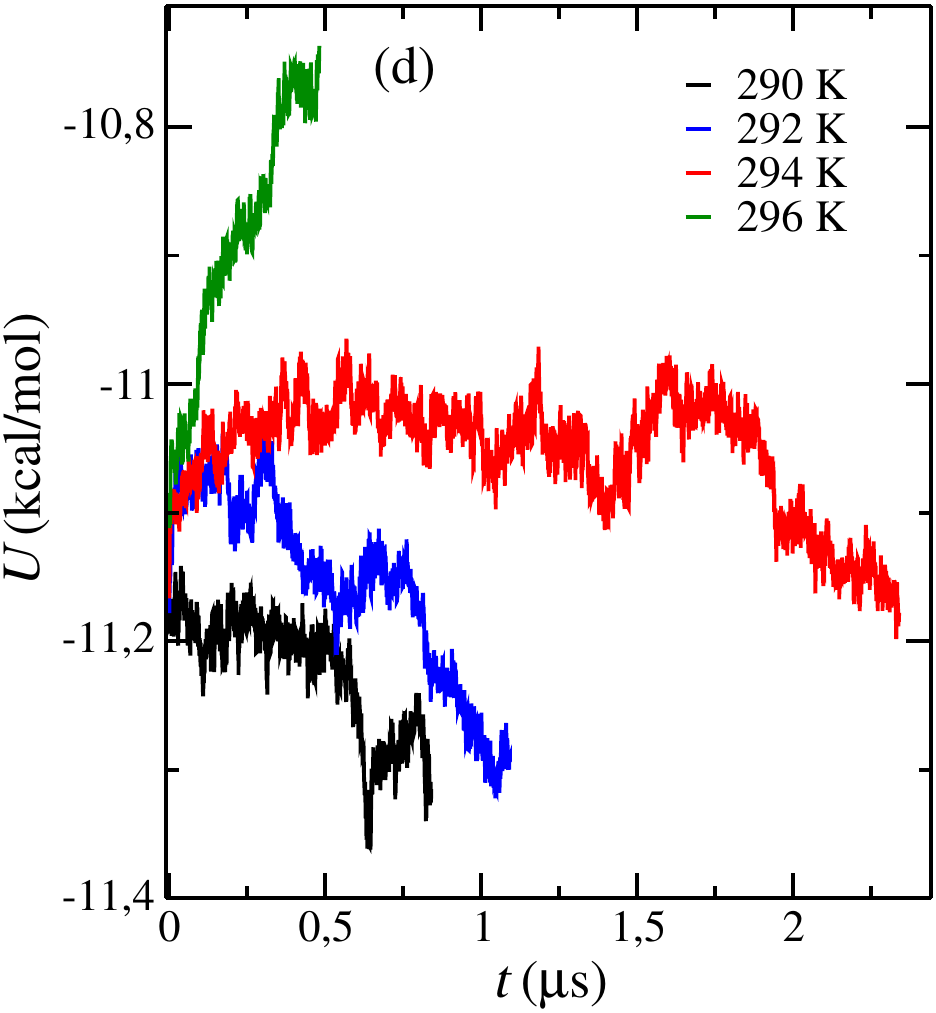}
         \includegraphics[width=0.30\textwidth]{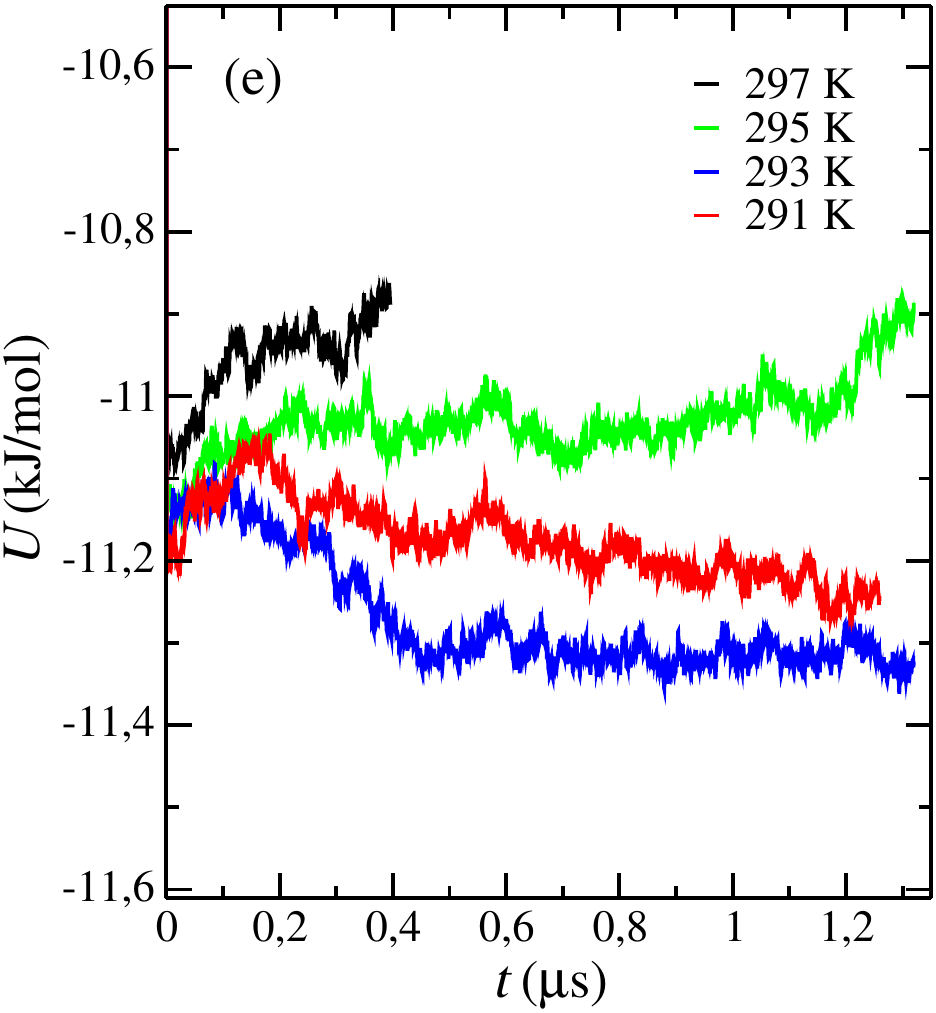}
         \includegraphics[width=0.30\textwidth]{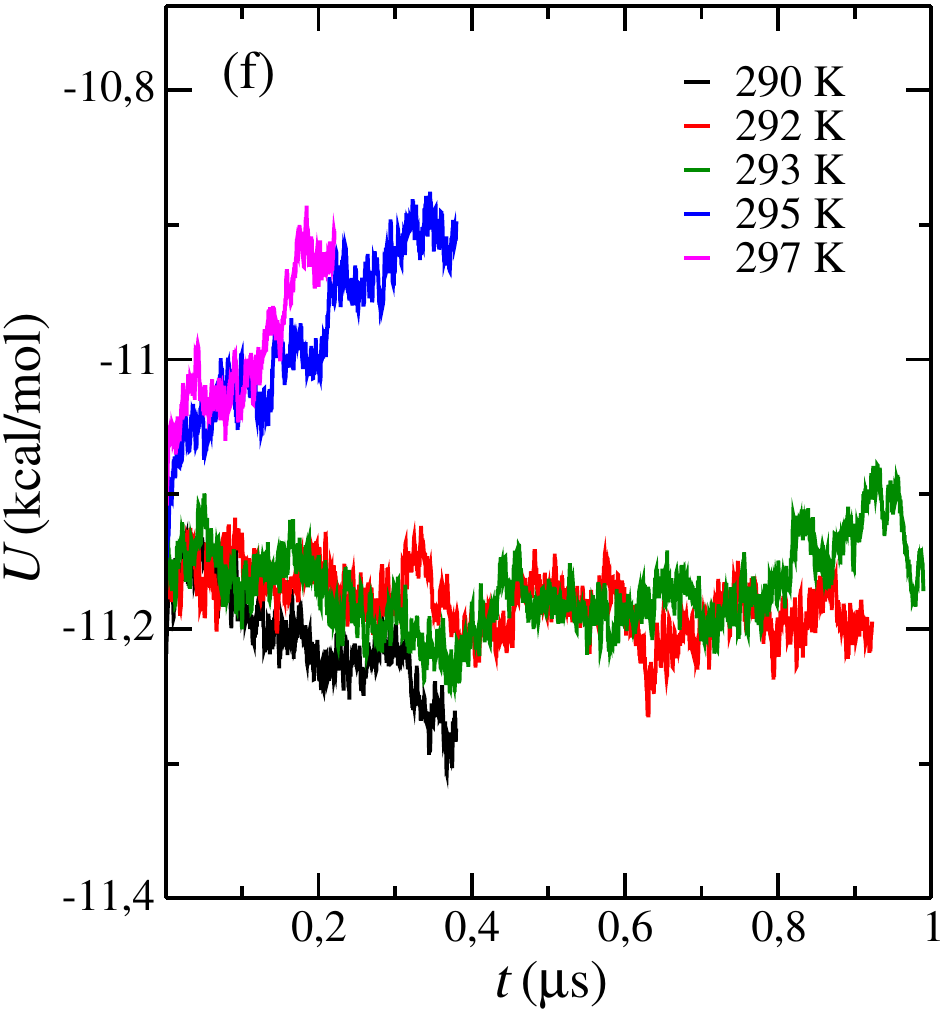}
        \caption{Potential energy, as a function of time, at $400\,\text{bar}$ and different temperatures for different cut-off values for the methane hydrate-water-gas methane system: (a) $0.9\,\text{nm}$, (b) $1.1\,\text{nm}$, (c) $1.3\,\text{nm}$, (d) $1.6\,\text{nm}$, (e) $0.9\,\text{nm}$ + LRCH, and (f) $0.9\,\text{nm}$ + LRCI.}
        \label{energies-methane}
\end{figure*}

It is interesting to recall here the main conclusion of the effect of dispersive interactions on the dissociation temperature of the CO$_{2}$ hydrate at different pressures. As can be seen in Fig.~\ref{CO2_result}, the effect of the cut-off distance on the $T_3$ value depends on pressure: its effect is very important at the lowest pressure considered and it decreases as the pressure is increased. Particularly, when pressure is high enough, there is no effect of the cut-off distance on the stability conditions of the hydrate. In terms of efficiency, this should be taken into account because the use of a small cut-off distance provides the same result as that using a large cut-off value with a decrease in the computational cost. If simulations are performed using a cut-off distance of $1.6\,\text{nm}$, then simulations using cut-off distances of $1.3$, $1.1$, and $0.9\,\text{nm}$ (with or without LRCH) are $1.5$, $2.1$, and $3$ faster, respectively. It is clear from this discussion that, not only the system size affect the dissociation temperature of a hydrate, as it is shown in our previous papers I and II,~\cite{paperI,paperII} but also the election of the cut-off distance strongly affects the dissociation temperature of hydrates.

Finally, an additional clarification is needed on the relation between the cut-off distance and the unlike intermolecular interactions for the CO$_{2}$ hydrate. Contrary to what happens with the CH$_{4}$ hydrate, in the case of the CO$_{2}$ hydrate it is necessary to fit the unlike water--carbon dioxide dispersive interaction to obtain a quantitative description of the dissociation temperature of this hydrate as it was shown several years ago by  M\'{\i}guez \emph{et al.}~\cite{Miguez2015a} and Costandy \emph{et al.}~\cite{Costandy2015a} The election of the optimal value of $\xi$ in both studies was done using a relatively small cut-off distance and system size. Particularly, both studies used $r_{c}=1.0$ and $0.9\,\text{nm}$, respectively, and a hydrate seed obtained replicating the unit cell 2 times along each space direction ($2\times 2\times 2$). Using these parameters, it was possible to provide an excellent agreement between simulation and experiments, particularly in the case of the study by M\'{\i}guez \emph{et al.}~\cite{Miguez2015a} In this work, however, we use larger cut-off distances and a larger hydrate seed ($3\times 3\times 3$). As a consequence of this, agreement between simulations and experimental data taken from the literature becomes worse as the cut-off distance increases. This is a direct consequence of fitting $\xi$ using a particular set of $r_{c}$ value and system size. If a better agreement between simulation and experiments is needed, a new modification of the Berthelot rule should be developed using larger cut-off distances and system sizes in order to match the experimental data.~\cite{Sloan2008a} Unfortunately, this is out of the scope of this work.

\subsection{CH$_4$ Hydrate}
We now examine the effect of the cut-off distance on the $T_{3}$ of the CH$_{4}$ hydrate. For this purpose, we have used the same system size employed for the CO$_2$ hydrate (see Table \ref{tabla-moleculas}). Note that methane hydrate usually requires larger simulation times (typically, simulation times are of the order of $\mu\text{s}$) than those used for the CO$_2$ hydrate to observe either growth or melt. This is due to the high solubility of CO$_2$ in water, about ten more times, compared with that of CH$_{4}$ in water. In Figure \ref{energies-methane}, we show the potential energy, as a function of time, of the studied system at $400\,\text{bar}$ and different temperatures using several cut-off distances. Similarly, we have used four distinct values, ranging from $0.9$ to $1.6\,\text{nm}$, as in the case of the CO$_{2}$ hydrate. Additionally, we have implemented two different LRCs with $r_{c}=0.9\,\text{nm}$. On one hand, we have incorporated the same LRCH corrections for dispersive energy and pressure than in the study of the CO$_{2}$ hydrate. It is important to note that these corrections can be only applied if the system is homogeneous. This is not the case of systems under study since we are simulating inhomogeneous systems in which gas, liquid, and solid phases coexist via planar interfaces. On the other hand, we have applied LRCs using the PME method 
for dispersive interactions, which is valid for inhomogeneous systems (LRCI).

The analysis of the potential energies, as functions of time, yields several notable findings. Firstly, it is evident that the time required for this hydrate to melt or grow is significantly longer compared to that of the CO$_{2}$ hydrate. In fact, we have conducted multiple simulations in which the CH$_{4}$ hydrate needs more than $2\,\mu$ to grow. Note that in the case of the CO$_{2}$ hydrate, the longest time observed for growing is only $1\,\mu$. This behavior aligns with both  experimental\cite{freer2001methane,touil2019gas,uchida1999microscopic,wells2021carbon,nagashima2020film,ou2016situ,daniel2015hydrate} and \textit{in silico}\cite{blazquez2023growth,tung2010growth} observations, which consistently show that the CH$_{4}$ hydrate exhibits larger growth rates compared to those of the CO$_{2}$ hydrate.

\begin{figure}[hbt!]
     \centering
         \includegraphics[width=\columnwidth]{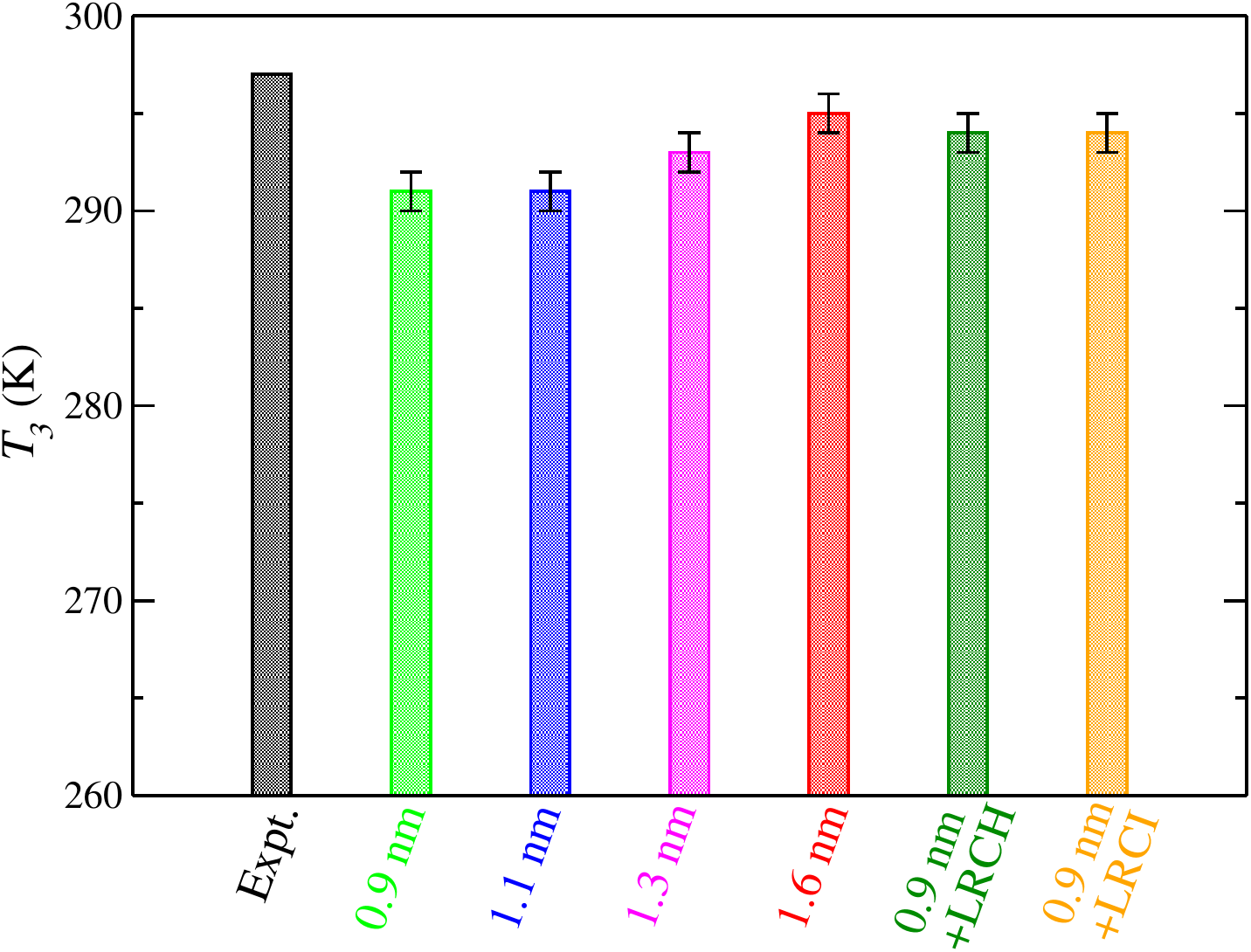}
        \caption{Dissociation temperature of the CH$_4$ hydrate, at $400\,\text{bar}$, as a function of the cut-off distance. Results obtained in this work are presented as colored bars and uncertainties as black lines.}
        \label{confs-cutoff}
\end{figure}

Secondly, the dissociation temperature $T_{3}$ varies with each cut-off distance used in simulations. Fig.~\ref{confs-cutoff} displays the $T_{3}$ obtained using different cut-off distances in a bar chart. It is evident that $T_{3}$ increases as the cut-off value is increased. This trend mirrors that observed in the case of the CO$_{2}$ hydrate, indicating a common behavior in both systems with respect to cut-off values. Notice that at the same pressure (400 bar) for the CO$_2$ the T$_3$ increases in 2 K when going from 0.9 nm to 1.1 nm in the cut-off, whereas in the case of methane we observe an increment of 1 K. It is neccesary to apply the LRCs in this system to observe the increment of 2 K. 

As in Section IV A, we have also performed simulations using a cut-off distance of $0.9\,\text{nm}$ along with LRCs to both energy and pressure. As can be seen, we also observe a notable increase in the $T_{3}$ value when compared to that obtained using $r_{c}=0.9\,\text{nm}$ without any LRCs. It is interesting to compare the predictions obtained using this value of cut-off in combination with homogeneous (LRCH) and inhomogeneous (LRCI) corrections. As can be seen, the $T_{3}$ obtained in both cases is identical and also very close to the value obtained with $r_{c}=1.6\,\text{nm}$. In fact, when considering the associated error bars, the dissociation temperatures are practically the same. 

To finish this section, it is worthy to mention the efficiency in terms of computational cost. According to Fig.~\ref{confs-cutoff}, simulations using a cut-off distance $r_{c}=0.9\,\text{nm}$, in combination with LRCs, and using a cut-off distance $r_{c}=1.6\,\text{nm}$ predict virtually identical dissociation temperatures within the error bars. It is important to take into account that simulations needed to distinguish if the hydrate grows or melts take $2\mu\text{s}$ of simulated time in most cases. According to this, we recommend conducting simulations using a cut-off distance equal to $0.9\,\text{nm}$ and the inclusion of LRCs for the case of the CH$_{4}$ hydrate. This approach not only conserves computational resources but also yields results that are consistent with the higher computational cost option, making it a more efficient and practical choice. The $T_3$ values obtained in this work for the CH$_4$ hydrate at 400 bar and each cut-off have been summarized in Table \ref{tabla-T3}.

\section{Conclusions}

In this work, we have employed the direct coexistence simulation technique to study the three-phase equilibria of CO$_{2}$ and CH$_{4}$ hydrates. The dissociation temperature, $T_{3}$, of both hydrates, has been previously studied by some of us.\cite{Conde2010a,Miguez2015a} Nevertheless, in this case, we have used a larger system to avoid finite-size effects\cite{Conde2017a}. By employing this system size for both hydrates, namely, CO$_{2}$ hydrate--aqueous solution--liquid CO$_{2}$ and CH$_{4}$ hydrate--aqueous solution--gas CH$_{4}$, we have performed simulations using different cut-off distances to elucidate its role in the determination of the $T_{3}$ of both hydrates.

Our initial investigation focuses on examining the impact of cut-off distance on the dissociation temperature of the CO$_{2}$ hydrate at various pressures. 
 In general, an increase of the cut-off distance provokes a shift of the $T_{3}$ values of the CH$_{4}$ and CO$_{2}$ hydrates. This is also true for all the pressures considered in the case of the CO$_{2}$ hydrate. This was expected since larger cut-off distances imply that more attractive interactions are taken into account, which results in a more stable hydrate structure. In other words, the $T_{3}$ is shifted towards higher temperatures. More subtle is how the $T_{3}$ is modified for different pressures. We observe that the influence of the cut-off distance on the $T_{3}$ is more significant at lower pressures. Specifically, at $100$ and $400\,\text{bar}$, increasing the cut-off value produces an increase in $T_3$. However, at $1000\,\text{bar}$, the effect of the cut-off on the dissociation temperature becomes negligible for $r_{c}\ge1.1\,\text{nm}$. This suggests that the effect of the cut-off value on $T_{3}$ is pressure-dependent, with a more pronounced effect at lower pressures.

Once we study the effect of the cut-off on the dissociation temperature of the CO$_{2}$ hydrate at different pressures, we consider a hydrate with a different guest. For its interesting applications, we select the CH$_{4}$ hydrate for this work. We concentrate only on one pressure, $400\,\text{bar}$, since simulation times required at lower temperatures are computationally expensive. In this case, we find the same effect observed for the CO$_2$ hydrate, i.e., an increase of the cut-off values yields a rise of the $T_{3}$ of the CH$_{4}$ hydrate. 

Based on the preceding findings, a pressing question arises: What cut-off value should be employed to accurately predict the \textit{T$_3$} for both hydrates? The use of larger cut-off distances provides better descriptions of the full potential model. The key point is to know which is the lowest value of $r_{c}$ able to capture the full potential (optimal value of $r_{c}$). Obviously, larger cut-off distance values than this optimal value would provide the same $T_{3}$ values but with a higher computational cost. Unfortunately, there is not a definitive answer to this query. Our observations reveal varying scenarios contingent upon pressure conditions.
For instance, in the case of the CO$_{2}$ hydrate at $100\,\text{bar}$, the most precise predictions are achieved with a $1.6\,\text{nm}$ cut-off value. However, at $400\,\text{bar}$, a smaller $1.3\,\text{nm}$ cut-off suffices. Finally, at $1000\,\text{bar}$, the use of $r_{c}=0.9\,\text{nm}$ and LRCs proves adequate for providing a reliable $T_{3}$ estimation. Conversely, using a different guest molecule in the hydrate, CH$_{4}$, the situation changes. In the case of methane at 400 bar, a $0.9\,\text{nm}$ cut-off distance and LRCs yields reasonably $T_{3}$ estimations. There exists no universally applicable answer to the previously mentioned question. The impact of the cut-off value is contingent on both the guest molecule and the prevailing pressure conditions. In essence, each system is different and necessitates a thorough investigation to determine the most suitable cut-off distance for the precise estimation of $T_{3}$ However, as we previsouly mentioned, we recommend simulating using the cut-off of 0.9 nm and applying LRCs. In this way, we are able to simulate the system with affordable computational resources and  to obtain accurate results for the T$_3$.

In summary, this work provides valuable insights into the intricate process of computing the dissociation line of hydrates. It suggests a specific cut-off value that can yield accurate estimations of $T_{3}$ for the examined force fields without necessitating excessively time-consuming simulations. This finding contributes to the optimization of computational resources while maintaining the reliability of the results in the study of hydrate systems.

\section*{Acknowledgments}

This work was funded by Ministerio de Ciencia e Innovaci\'on (Grant No.PID2019-105898GA-C22, PID2021-125081NB-I00 and PID2022-136919NB-C32), Junta de Andalucía (P20-00363), and Universidad de Huelva (P.O. FEDER UHU-1255522 and FEDER-UHU-202034), all four cofinanced by EU FEDER funds. This work was also funded by Project No.~ETSII-UPM20-PU01 from ``Ayudas Primeros Proyectos de la ETSII-UPM''. M.M.C. acknowledges CAM and UPM for financial support of this work through the CavItieS project No. APOYO-JOVENES-01HQ1S-129-B5E4MM from ``Accion financiada por la Comunidad de Madrid en el marco del Convenio Plurianual con la Universidad Politecnica de Madrid en la linea de actuacion estimulo a la investigacion de jovenes doctores'' and CAM under the Multiannual Agreement with UPM in the line Excellence Programme for University Professors, in the context of the V PRICIT (Regional Programme of Research and Technological Innovation). S.B. acknowledges Ayuntamiento de Madrid for a Residencia de Estudiantes grant. The authors also gratefully acknowledge the Universidad Politecnica de Madrid (www.upm.es) for providing computing resources on Magerit Supercomputer. We also acknowledge additional computational resources from Centro de Supercomputaci\'on de Galicia (CESGA, Santiago de Compostela, Spain), at which some of the simulations were run.

\section*{AUTHORS DECLARATIONS}

\section*{Conflicts of interest}

The authors have no conflicts to disclose.

\section*{Data availability}

The data that support the findings of this study are available within the article.

\section*{References}
\bibliography{bibfjblas}

\end{document}